\documentclass[11pt]{article}
\usepackage{epsfig}

\makeatletter\@addtoreset{equation}{section}\makeatother
\def\ket#1{\left| #1 \right\rangle}
\def\vev#1{\left\langle #1 \right\rangle}
\def\no#1{{\rm :} \,  #1 \, {\rm :}}
\renewcommand{\varepsilon}{\lambda}

\begin{document}
\baselineskip=15.5pt
\renewcommand{\theequation}{\arabic{section}.\arabic{equation}}
\pagestyle{plain}
\setcounter{page}{1}
\def\appendix{
\par
\setcounter{section}{0}
\setcounter{subsection}{0}
\def\thesection{\Alph{section}}}

\begin{titlepage}
\rightline{\tt hep-th/0503068}
\rightline{\tt MIT-CTP-3606}
\begin{center}
\vskip 2.5cm
{\Large \bf {Vacuum string field theory
without matter-ghost factorization}}
\vskip 1.0cm
{\large {Nadav Drukker${}^1$ and Yuji Okawa${}^2$}}
\vskip 1.0cm
{\it {${}^1$ The Niels Bohr Institute, Copenhagen University}}\\
{\it {Blegdamsvej 17, DK-2100 Copenhagen, Denmark}}\\
drukker@nbi.dk
\vskip 0.5cm
{\it {${}^2$ Center for Theoretical Physics, Room 6-304}}\\
{\it {Massachusetts Institute of Technology}}\\
{\it {Cambridge, MA 02139, USA}}\\
okawa@lns.mit.edu
\vskip 1.8cm
{\bf Abstract}
\end{center}

\noindent
We show that
vacuum string field theory with the singular kinetic operator
conjectured by Gaiotto, Rastelli, Sen and Zwiebach
can be obtained by field redefinition from a regular theory
constructed by Takahashi and Tanimoto.
We solve the equation of motion
both by level truncation
and by a series expansion using the regulated butterfly state,
and we find evidence that the energy density of a D25-brane
is well defined and finite.
Although the equation of motion naively factorizes
into the matter and ghost sectors in the singular limit,
subleading terms in the kinetic operator are relevant
and the factorization does not strictly hold.
Nevertheless,
solutions corresponding to different D-branes can be constructed
by changing the boundary condition
in the boundary conformal field theory formulation
of string field theory,
and ratios of D-brane tensions are shown to be reproduced correctly.

\end{titlepage}

\newpage

\section{Introduction}

Vacuum string field theory
\cite{Rastelli:2000hv, Rastelli:2001jb, Rastelli:2001uv}
is a conjectured form of
Witten's cubic open string field theory~\cite{Witten:1985cc}
expanded around the tachyon vacuum.\footnote
{
For a recent review, see \cite{Taylor:2003gn}.
}
The action of vacuum string field theory
is given by replacing the BRST operator
in Witten's string field theory
with a different operator which we denote by ${\cal Q}$.
It was conjectured in \cite{Rastelli:2000hv} that
this operator can be made purely of ghost fields,
and under this assumption the equation of motion
factorizes into the matter and ghost sectors.
Then the matter part of the equation can be solved
by the matter sector of a star-algebra projector
\cite{Rastelli:2001jb, Kostelecky:2000hz}
such as the sliver state \cite{Rastelli:2001vb, Rastelli:2000iu}
or the butterfly state \cite{Gaiotto:2001ji, Schnabl:2002ff,
Gaiotto:2002kf, Okawa:2003cm}.
Using the conformal field theory (CFT) formulation
of string field theory \cite{LeClair:1988sp, LeClair:1988sj},
we can construct such a star-algebra projector
for any given consistent open-string boundary condition
\cite{Rastelli:2001vb}.
The resulting solution is conjectured to describe the D-brane
corresponding to the open-string boundary condition
\cite{Rastelli:2001vb, Mukhopadhyay:2001ey}.
Based on this description of D-branes,
it has been shown that ratios of D-brane tensions
\cite{Rastelli:2001vb, Okuyama:2002tw},
the open-string mass spectrum on any D-brane \cite{Okawa:2002pd},
and the absolute value of the D25-brane tension \cite{Okawa:2002pd}
can be reproduced correctly.
The factorization into the matter and ghost sectors
is a key ingredient in deriving these results.

A more specific conjecture on the operator ${\cal Q}$ 
was put forward later by Gaiotto, Rastelli, Sen and Zwiebach 
\cite{Gaiotto:2001ji}.
It was conjectured that the kinetic operator
is given by a singular limit
where ${\cal Q}$ becomes
a $c$-ghost insertion at the open-string midpoint
with a divergent coefficient.
With this kinetic operator,
the full equation of motion including the ghost sector
can be solved by a star-algebra projector
in the twisted CFT \cite{Hata:2001sq, Gaiotto:2001ji}
or by an ordinary star-algebra projector
with an appropriate $c$-ghost insertion~\cite{Okawa:2003zc}.

There are, however, remaining problems.
It was shown in \cite{Okawa:2003cm, Okawa:2003zc} that
when the solution $\ket{\Psi}$ is given by
the twisted butterfly state or by the butterfly state
with a $c$-ghost insertion,
the equation of motion contracted with the solution $\ket{\Psi}$,
\begin{equation}
  \langle \Psi | {\cal Q} | \Psi \rangle
  + \langle \Psi | \Psi \ast \Psi \rangle = 0 \,,
\label{equation-contracted-with-Psi}
\end{equation}
turns out to be incompatible with the equation contracted
with a state $\ket{\phi}$ in the Fock space:
\begin{equation}
  \langle \phi | {\cal Q} | \Psi \rangle
  + \langle \phi | \Psi \ast \Psi \rangle = 0 \,.
\label{equation-contracted-with-phi}
\end{equation}
This follows from the fact that the ratio
\begin{equation}
  \frac{\langle \phi | \Psi \ast \Psi \rangle}
       {\langle \phi | {\cal Q} | \Psi \rangle}
  \frac{\langle \Psi | {\cal Q} | \Psi \rangle}
       {\langle \Psi | \Psi \ast \Psi \rangle}
\label{compatibility}
\end{equation}
is different from $1$ in the limit
where ${\cal Q}$ becomes
the $c$-ghost insertion at the open-string midpoint.
Since this ratio is independent of the normalizations
of ${\cal Q}$, $\ket{\Psi}$, and $\ket{\phi}$,
the conclusion holds whatever scaling limit
we may take.
One resolution suggested in \cite{Okawa:2003cm}
is the possible relevance of the subleading terms of ${\cal Q}$.
Since the solution $\ket{\Psi}$ is outside the Fock space,
the subleading terms in ${\cal Q}$ can contribute
to $\langle \phi | {\cal Q} | \Psi \rangle$
or to $\langle \Psi | {\cal Q} | \Psi \rangle$
at the same order as the leading term.
In fact, it was demonstrated using an explicit example
in \cite{Okawa:2003cm} that this is indeed possible.

Another issue is the finiteness of the physical coupling constant.
Since vacuum string field theory is formulated around a vacuum
without physical open string states, the coupling constant
in the action is not directly related to physical quantities.
The string coupling constant can be read off, for example, from
a three-tachyon scattering amplitude around a D25-brane solution
as in \cite{Okawa:2002pd}.
The finiteness of the string coupling constant requires
the existence of a D-brane solution with finite energy density.
In the case of the twisted butterfly state solution,
it was shown in \cite{Okawa:2003cm}
that the energy density can be finite in the limit
if one appropriately scales the parameter of the solution
as the coefficient
of the singular pure-ghost kinetic operator diverges.
However, whether or not this scaling limit is realized
cannot be determined
by the analysis at the leading order
and depends on the subleading structure
of the theory \cite{Okawa:2003cm}.
This also indicates the relevance of the subleading terms
in the kinetic operator.

If the subleading terms in ${\cal Q}$ are relevant,
the factorization into the matter and ghost sectors
at the leading order may be ruined.
Since the factorization was a key assumption
in deriving the various results mentioned earlier,
this could be a serious problem.
For example, the construction of other D-brane solutions
with correct tension ratios may be no longer valid.
Constructing a consistent next-to-leading term
in the kinetic operator,
however, is a nontrivial problem,
which has prevented us from addressing this issue
in a specific example.

In this paper, we claim that
the assumption of the matter-ghost factorization
in the kinetic operator ${\cal Q}$
can be relaxed
in constructing D-brane solutions
with correct tension ratios.
We demonstrate this in a concrete setting
based on the regularization
of the singular pure-ghost kinetic operator
proposed in \cite{Drukker:2003hh}.
We first show in section \ref{section-2} that
vacuum string field theory with the singular kinetic operator
conjectured by Gaiotto, Rastelli, Sen and Zwiebach
can be realized by field redefinition
of a regular theory constructed
by Takahashi and Tanimoto \cite{Takahashi:2002ez}.
Then in section \ref{section-3},
we solve the equation of motion
both by level truncation
and by a series expansion
using the regulated butterfly state \cite{Okawa:2003zc},
and we find evidence that the energy density of a D25-brane
is well defined and finite in this theory.
The long-standing issue of the finiteness
of the physical coupling constant therefore
seems to be resolved in this particular realization
of vacuum string field theory
with the singular pure-ghost kinetic operator.
In this case, however, the subleading terms
in the kinetic operator are in fact relevant,
and their contributions to physical quantities
such as the D25-brane tension
are nonvanishing in the singular limit.
Therefore, the apparent factorization
into the matter and ghost sectors does not strictly hold.
Nevertheless, we show in section \ref{section-4} that
solutions corresponding to different D-branes can be constructed
by changing the boundary condition in the boundary CFT formulation,
and ratios of D-brane tensions can be reproduced correctly.
This argument in section \ref{section-4} is general
and is not restricted to this particular choice
of the kinetic operator.
In fact, we do not even need to assume that
the kinetic term can be brought to the singular
$c$-ghost insertion at the open-string midpoint.
We comment on our results and discuss open problems
in section \ref{section-5}.

We use standard conventions
in the CFT formulation of string field theory,
which are summarized in appendix A.
Some details of subsection \ref{subsection-3.2}
are presented in appendix B.

\section{Singular pure-ghost theory from a regular theory}
\label{section-2}

\subsection{Singular pure-ghost vacuum string field theory}
\label{subsection-2.1}

The action of Witten's cubic open string field theory
\cite{Witten:1985cc} is given by
\begin{equation}
  S = -\frac{1}{\alpha'^3 g_T^2} \left[
      \frac{1}{2} \langle \Psi | Q_B | \Psi \rangle
      + \frac{1}{3} \langle \Psi | \Psi \ast \Psi \rangle \right],
\end{equation}
where $g_T$ is the on-shell three-tachyon coupling constant.
The action of vacuum string field theory is derived from this action
by expanding it around the solution
corresponding to the tachyon vacuum,
and it will take the same form except for the kinetic operator.
It was conjectured in \cite{Rastelli:2000hv}
that the resulting kinetic operator can be made purely
of ghost fields by field redefinition.
We write the action of vacuum string field theory as
\begin{equation}
  S = - \frac{1}{\alpha'^3 g^2} \left[
  \frac{1}{2} \langle \Psi | {\cal Q} | \Psi \rangle
  + \frac{1}{3} \langle \Psi | \Psi \ast \Psi \rangle \right],
\end{equation}
where we used a different notation $g$ for the coupling constant
because the field redefinition may change it from the original
value $g_T$.
We will discuss this point further
in subsection \ref{subsection-2.3}.
The equation of motion derived from this action
factorizes into the matter and ghost sectors
if ${\cal Q}$ can be made purely of ghost fields.

A more specific conjecture on ${\cal Q}$ was put forward later
in \cite{Gaiotto:2001ji}.
It was conjectured that there exists a one-parameter family
of field redefinition which takes ${\cal Q}$ to the following
form:
\begin{equation}
  {\cal Q} = \frac{Q}{2 \epsilon} \left[ 1 + o(\epsilon) \right] \,,
\label{Q-conjecture}
\end{equation}
where $Q$ is a midpoint $c$-ghost insertion
defined by
\begin{equation}
  Q = \frac{1}{2i} (c(i) - c(-i)) \,,
\end{equation}
and $\epsilon$ is
the parameter of the field redefinition.
We denoted terms which vanish in the limit $\epsilon \to 0$
by $o(\epsilon)$, which does not have to be $O(\epsilon)$.
In the singular limit $\epsilon \to 0$, the midpoint $c$-ghost
insertion $Q$ dominates in the kinetic operator ${\cal Q}$
with an infinite coefficient.

If there are no subleading terms in (\ref{Q-conjecture}),
the parameter $g$ can be absorbed
by field redefinition~\cite{Okawa:2003cm}.
Therefore, the subleading terms are necessary
in order for the theory to have
a parameter corresponding to the string coupling constant.\footnote
{
A different interpretation was suggested
in \cite{Bonora:2003cc, Bonora:2004qj}.}

Since vacuum string field theory is formulated around a vacuum
where there are no physical open strings,
the coupling constant $g$ in the action is not directly
related to physical quantities.
The physical string coupling constant can be calibrated,
for example, by the energy density ${\cal E}$ of a D25-brane.
Since ${\cal E} = 1/(2 \pi^2 \alpha'^3 g_T^2)$,
$g_T$ is given by
\begin{equation}
  g_T = \frac{1}{\sqrt{2 \pi^2 \alpha'^3 {\cal E}}} \,.
\end{equation}
This shows that the energy density of the D25-brane solution
in vacuum string field theory should be finite
in order to have a finite string coupling constant.
It would be less interesting if we only formulated
a theory with a vanishing or divergent string coupling constant.

There are other ways to identify
the string coupling constant
in vacuum string field theory.
For example, it can be identified by
the on-shell three-tachyon coupling
around a D25-brane solution.
Strictly speaking, $\alpha'$ should also be identified,
for example, from the mass spectrum of open strings
around a D25-brane.
It is a nontrivial check if the energy density
of the D25-brane solution can be written as
$1/(2 \pi^2 \alpha'^3 g_T^2)$
with the correct numerical coefficient
in terms of $\alpha'$ and $g_T$ identified in this way.
It was shown in \cite{Okawa:2002pd}
that this expression  can be in fact reproduced
in vacuum string field theory
under the assumption of the matter-ghost factorization.

The equation of motion
with ${\cal Q}$ given by (\ref{Q-conjecture})
can be solved by a star-algebra projector in the twisted CFT
in the limit $\epsilon \to 0$ \cite{Gaiotto:2001ji}.
It was shown in \cite{Okawa:2003cm} that
the energy density of the solution given by
the twisted butterfly state can be finite
if one appropriately scales the regularization parameter
of the twisted butterfly state in the limit $\epsilon \to 0$.
However, whether or not this scaling is realized
cannot be determined from the analysis at the leading order,
and it depends on the subleading terms \cite{Okawa:2003cm}.

To summarize, the subleading terms in (\ref{Q-conjecture})
are necessary in order for the theory to have a parameter
corresponding to the string coupling constant.
Furthermore, the finiteness of the physical coupling constant
depends on the subleading terms in (\ref{Q-conjecture}).

\subsection{Regularizing the singular pure-ghost operator}
\label{subsection-2.2}

To address these issues 
it is therefore important to regularize
the singular pure-ghost kinetic operator~(\ref{Q-conjecture}).
However, it is a nontrivial problem to construct
a next-to-leading term in (\ref{Q-conjecture})
which satisfies the conditions for a consistent kinetic operator
of string field theory such as ${\cal Q}^2 = 0$.
One way to regularize the singular pure-ghost operator
was presented in \cite{Drukker:2003hh}
by modifying a class of kinetic operators
constructed by Takahashi and Tanimoto \cite{Takahashi:2002ez}.

The operators constructed in \cite{Takahashi:2002ez}
take the following form:
\begin{equation}
  Q_f = \oint_C \frac{d \xi}{2\pi i}
  \biggl[ f(\xi) \, j_B(\xi) - \frac{f'(\xi)^2}{f(\xi)} \, c(\xi)
  \biggr] \,,
\label{Q_f-definition}
\end{equation}
where $j_B(\xi)$ is the BRST current
and $f(\xi)$ is any analytic function
satisfying $f(-1/\xi)=f(\xi)$.
The contour $C$ runs on the unit circle counterclockwise.
The operator $Q_f$ is formally related to $Q_B$
by a similarity transformation
so one may expect the theory with $Q_f$ to be equivalent
to the theory with $Q_B$.
However, the similarity transformation can be singular,
and the theory with $Q_f$ can be inequivalent
to the theory with $Q_B$.
The simplest such example considered in \cite{Takahashi:2002ez}
is $f(\xi) = -(1-\xi^2)^2/(4 \xi^2)$.
It was further required in \cite{Takahashi:2002ez}
that $f(\xi)$ is analytic inside the unit disk
except for a pole at the origin.
However, $Q_f$ can be a consistent kinetic operator
of string field theory even if $f(\xi)$ has poles
away from the origin, and a more general class
of operators were discussed in \cite{Drukker:2003hh}.\footnote
{
After finishing the paper, we learned that the authors of
\cite{Kishimoto:2002xi} were aware of this generalization
of the operators constructed in \cite{Takahashi:2002ez}.
}
In particular, if we choose $f(\xi)$ to be
\begin{equation}
  f(\xi) = \frac{\epsilon^2 \left( 1 - \xi^2 \right)^2}
           {\epsilon^2 \left( 1 - \xi^2 \right)^2
           -\left( 1 + \xi^2 \right)^2} \,,
\label{f(xi)}
\end{equation}
$Q_f$ takes
the form of (\ref{Q-conjecture}) \cite{Drukker:2003hh}.
Therefore, we can use this operator
to regularize the singular pure-ghost operator.
Note, however, that this class of operators are not
made purely of ghost fields and depend on the matter sector
through the BRST current $j_B (\xi)$.
Therefore, the factorization into the matter and ghost sectors
does not hold for finite $\epsilon$.
The function $f(\xi)$ has double zeros at $\xi = \pm 1$,
which play an important role later in section \ref{section-4}.
When $\epsilon < 1$,
the poles of $f(\xi)$ are located along the imaginary axis at
\begin{equation}
  \xi = \pm i \sqrt{\frac{1-\epsilon}{1+\epsilon}} \,, \qquad
        \pm i \sqrt{\frac{1+\epsilon}{1-\epsilon}} \,.
\end{equation}
They become two double poles at the origin and at infinity
when $\epsilon = 1$,
and they dissolve into single poles on the real axis
when $\epsilon > 1$.
The contour $C$ cannot be deformed across any of these poles.

Let us first verify that $Q_f^2$ vanishes. From
the operator product expansions (OPE's),
\begin{eqnarray}
  j_B(z) \, j_B(w)
  &\sim& - \frac{4}{(z-w)^3} c\partial c(w)
  - \frac{2}{(z-w)^2} c\partial^2 c(w)
  = -\partial_w \left[ \, \frac{2}{(z-w)^2} c\partial c(w)
  \, \right] \,,
\nonumber\\*
  j_B(z) \, c(w) &\sim& \frac{1}{z-w} c\partial c(w)\,,
\end{eqnarray}
it follows that
the commutator involving two BRST currents
cancels the cross terms involving $j_B$ and $c$.
The necessary deformation of the integration contours
can be made without crossing the poles.

Next we verify that $Q_f$ is a derivation of the star product:
\begin{equation}
  Q_f | A \ast B \rangle
  = | Q_f A \ast B \rangle
  + (-1)^{\epsilon (A)} | A \ast  Q_f B \rangle \,,
\end{equation}
where $A$ and $B$ are string fields,
and $\epsilon (A) = 0$ mod 2 when $A$ is Grassmann even,
and $\epsilon (A) = 1$ mod 2 when $A$ is Grassmann odd.
If we write $Q_f$ in terms of modes defined by
\begin{equation}
  c_n = \oint_C \frac{d \xi}{2 \pi i} \, \xi^{n-2} \, c (\xi) \,,
\qquad
  Q_n = \oint_C \frac{d \xi}{2 \pi i} \, \xi^n \, j_B (\xi) \,,
\end{equation}
it takes the form of a linear combination
of $c_n + (-1)^n c_{-n}$ and $Q_n + (-1)^n Q_{-n}$
because of the condition $f(-1/\xi) = f(\xi)$.
We can show that $c_n + (-1)^n c_{-n}$ and $Q_n + (-1)^n Q_{-n}$
are derivations using the method in \cite{Rastelli:2000iu},
and this formally proves that $Q_f$ is a derivation.
Although there is a potential subtlety in this argument
because $Q_f$ has infinitely many terms
and $f(\xi)$ has poles inside the unit disk
away from the origin,
we can verify that the necessary deformation
of integration contours can be made
without crossing the poles.\footnote
{In defining the three-string vertex,
we can use conformal transformations $h^{(1)} (\xi)$, $h^{(2)} (\xi)$,
and $h^{(3)} (\xi)$ given by
\begin{equation}
  h^{(1)} (\xi) = \tan \left( - \frac{\pi}{6}
  + \frac{2}{3} \arctan \xi \right) \,, \qquad
  h^{(2)} (\xi) = \tan \left( \frac{\pi}{6}
  + \frac{2}{3} \arctan \xi \right) \,, \qquad
  h^{(3)} (\xi)
  = - \cot \left( \frac{2}{3} \arctan \xi \right) \,.
\end{equation}
Then the poles of $f$ on the upper-half plane when $\epsilon < 1$
are located at
\begin{equation}
  i \coth \left( \frac{2}{3} \, {\rm arctanh} \,
  \sqrt{\frac{1-\epsilon}{1+\epsilon}} \, \right) \,, \qquad
  \tan \left( \pm \frac{\pi}{6} + \frac{2 i}{3} \, {\rm arctanh} \,
  \sqrt{\frac{1-\epsilon}{1+\epsilon}} \right) \,.
\end{equation}
We can explicitly verify that the necessary deformations
of the contours can be made without crossing these poles.}

We have confirmed that the operator $Q_f$ with $f(\xi)$
given by (\ref{f(xi)}) is qualified to be
a consistent kinetic operator of string field theory.
Let us now consider the limit $\epsilon \to 0$
and show that $Q_f$ takes the form of (\ref{Q-conjecture}).
In what follows, we write the dependence of $Q_f$ on $\epsilon$
explicitly as $Q_f (\epsilon)$.
It is convenient to rewrite $f(\xi)$ as
\begin{equation}
  f(\xi) = - \frac{\epsilon^2}{1-\epsilon^2}
  \frac{\left( \xi^2 - p^2
  - \displaystyle{\frac{2}{1+\epsilon}} \right)^2}
  { (\xi^2-p^2) \left( \xi^2 - p^2
  + \displaystyle{\frac{\strut 4 \, \epsilon}{1-\epsilon^2}}
  \right)} \,,
\label{f-of-xi}
\end{equation}
where
\begin{equation}
  p = i \sqrt{\frac{1-\epsilon}{1+\epsilon}} \,.
\end{equation}
The poles inside the contour are located at $\pm p$.
The functions $f(\xi)$ and $-f'(\xi)^2/f(\xi)$ can be expanded
about $\xi = \pm p$ as follows:
\begin{eqnarray}
  f(\xi) &=& {} \mp \frac{\epsilon}{2 \, p \, (1+\epsilon)^2}
             \frac{1}{\xi \mp p} + O ( (\xi \mp p)^0 ) \,,
\label{f-of-xi-expansion}
\\
  - \frac{f'(\xi)^2}{f(\xi)}
  &=& {} \pm \frac{\epsilon}{2 \, p \, (1+\epsilon)^2}
          \frac{1}{(\xi \mp p)^3}
  + \frac{p^2 + \epsilon \, (1 + 4 \, p^2) + 3 \, \epsilon^2 \, p^2}
         {4 \, p^2 \, (1+\epsilon)^2}
    \frac{1}{(\xi \mp p)^2}
\nonumber \\ &&
    {} \mp \frac{p^4 - \epsilon \, (p^2 - 2 \, p^4)
              + \epsilon^2 (1 - 4 \, p^2 - 2 \, p^4)
              - 3 \, \epsilon^3 (p^2 + 2 \, p^4)
              - 3 \, \epsilon^4 \, p^4}
           {4 \, \epsilon \, p^3 (1+\epsilon)^2} \frac{1}{\xi \mp p}
\nonumber \\ &&
    {} + O ( (\xi \mp p)^0 ) \,.
\label{f'^2/f-of-xi-expansion}
\end{eqnarray}
Therefore, $Q_f (\epsilon)$ is given by
\begin{eqnarray}
  Q_f (\epsilon) &=& - \frac{\epsilon}{2 \, p \, (1+\epsilon)^2}
             ( j_B (p) - j_B (-p) )
        + \oint_{C_0} \frac{d \xi}{2 \pi i} \, f(\xi) \, j_B (\xi)
\nonumber \\ && \quad
        + \frac{\epsilon}{4 \, p \, (1+\epsilon)^2}
        ( \partial^2 c (p) - \partial^2 c(-p) )
  + \frac{p^2 + \epsilon \, (1 + 4 \, p^2) + 3 \, \epsilon^2 \, p^2}
         {4 \, p^2 \, (1+\epsilon)^2}
        ( \partial c (p) + \partial c(-p) )
\nonumber \\ && \quad
  - \frac{p^4 - \epsilon \, (p^2 - 2 \, p^4)
              + \epsilon^2 (1 - 4 \, p^2 - 2 \, p^4)
              - 3 \, \epsilon^3 (p^2 + 2 \, p^4)
              - 3 \, \epsilon^4 \, p^4}
             {4 \, \epsilon \, p^3 (1+\epsilon)^2}
        ( c (p) - c(-p) )
\nonumber \\ && \quad
  - \oint_{C_0} \frac{d \xi}{2 \pi i} \,
  \frac{f'(\xi)^2}{f(\xi)} \, c(\xi) \,,
\label{Q_f-expansion}
\end{eqnarray}
where the contour $C_0$ encircles the origin counterclockwise.
The operator $Q_f (\epsilon)$ is now represented
as various operator insertions at $\xi = \pm p$
and integrals around the origin.
It is easy to show that
the coefficients of $f(\xi)$ and $f'(\xi)^2/f(\xi)$
in the Laurent expansion around the origin at any finite order
are of $O(\epsilon^2)$.
Therefore, in the limit $\epsilon \to 0$ the integrals vanish,
and $Q_f (\epsilon)$ is dominated by the insertion $c(p)-c(-p)$
with $p \to i$:
\begin{equation}
  Q_f (\epsilon) = \frac{Q}{2 \epsilon} + O(\epsilon) \,.
\label{Q-singular-limit}
\end{equation}
This takes the form of (\ref{Q-conjecture}).
Note that terms of $O(\epsilon^0)$ cancel
so that the next-to-leading order is $O(\epsilon)$.

To be more precise, the expansion (\ref{Q-singular-limit})
should be understood as
\begin{equation}
  \langle \phi_1 | Q_f (\epsilon) | \phi_2 \rangle
  = \frac{1}{2 \epsilon} \langle \phi_1 | Q | \phi_2 \rangle
  + O(\epsilon) \,,
\end{equation}
where $\ket{\phi_1}$ and $\ket{\phi_2}$ are states in the Fock space.
The expansion (\ref{Q-singular-limit}) does not necessarily hold
when states outside the Fock space are involved.
We therefore should not expand $Q_f (\epsilon)$
in terms of $\epsilon$ at intermediate steps of any computation.
Let us take the computation
of $\{ Q_f (\epsilon), Q \}$ as an example to explain this subtlety.
The anticommutator can be easily computed using the OPE
between $j_B$ and $c$, and the result is
\begin{equation}
  \{ Q_f (\epsilon), Q \} = \frac{1}{2i}
  \left( c \partial c (i) - c \partial c (-i) \right) \,.
\end{equation}
It can also be computed in the operator method.
Obviously, only the $j_B$ part of $Q_f (\epsilon)$ is relevant.
Since it is given by
\begin{equation}
  \oint_C \frac{d \xi}{2\pi i} \,
  f(\xi) \, j_B(\xi)
  = \frac{\epsilon}{1+\epsilon} \, Q_B
  + \frac{\epsilon}{(1+\epsilon)^2} \sum_{n=1}^{\infty}
  (-1)^n \left( \frac{1-\epsilon}{1+\epsilon} \right)^{n-1}
  ( Q_{2n} + Q_{-2n} ) \,,
\label{Q_n-coefficients}
\end{equation}
we find
\begin{eqnarray}
  \{ Q_f (\epsilon), Q \}
  &=& \frac{\epsilon}{1+\epsilon} \, \{ Q_B, Q \}
  + \frac{\epsilon}{(1+\epsilon)^2} \sum_{n=1}^{\infty}
  (-1)^n \left( \frac{1-\epsilon}{1+\epsilon} \right)^{n-1}
  \{ Q_{2n} + Q_{-2n}, Q \}
\nonumber \\
  &=& \left[ \frac{\epsilon}{1+\epsilon}
  + \frac{2 \epsilon}{(1+\epsilon)^2} \sum_{n=1}^{\infty}
  \left( \frac{1-\epsilon}{1+\epsilon} \right)^{n-1}\right]
  \{ Q_B, Q \}
\nonumber \\
  &=& \{ Q_B, Q \} = \frac{1}{2i}
  \left( c \partial c (i) - c \partial c (-i) \right) \,,
\end{eqnarray}
where we have used
$\{ Q_{2n}, Q \} = (-1)^n \{ Q_B, Q \}$,
which follows from the anticommutation relation
$\{ Q_n, c_m \} = \{ Q_B, c_{n+m} \}$.
Note that each term in the second line is of $O(\epsilon)$,
but the infinite sum gives the result of $O(1)$.
This demonstrates that we should not expand $Q_f (\epsilon)$
in terms of $\epsilon$ in this computation
because the operator $Q$ takes a state out of the Fock space.

\subsection{Field redefinition}
\label{subsection-2.3}

We have shown that $Q_f (\epsilon)$ defines
a one-parameter family of consistent string field theory actions.
In fact, the actions with different values of $\epsilon$
are related by field redefinition and are equivalent.
Let us introduce a different parameterization
through $\epsilon = e^{-4 a}$.
It can be easily shown that the field redefinition
generated by $K_2 = L_2 - L_{-2}$ changes
the value of the parameter:
\begin{equation}
  e^{b K_2} \, Q_f (e^{-4 a}) \, e^{-b K_2}
  = Q_f (e^{-4 (a+b)}) \,.
\label{similar}
\end{equation}
Since the exponentiation is straightforward,
it is sufficient to prove this identity to linear order in $b$.
The function $f(\xi)$ with $\epsilon = e^{-4(a+b)}$
is expanded in $b$ as follows:
\begin{eqnarray}
  f(\xi) &=& \frac{e^{-8 \, (a+b)} (1-\xi^2)^2}
  {e^{-8 \, (a+b)} (1-\xi^2)^2 - (1+\xi^2)^2}
\nonumber \\
  &=& \frac{e^{-8 \, a} (1-\xi^2)^2}
  {e^{-8 \, a} (1-\xi^2)^2 - (1+\xi^2)^2}
  -b \, \left( \xi^3 - \frac{1}{\xi} \right)
  \frac{\partial}{\partial \xi}
  \left[ \frac{e^{-8 \, a} (1-\xi^2)^2}
  {e^{-8 \, a} (1-\xi^2)^2 - (1+\xi^2)^2} \right] + O(b^2) \,. \quad
\end{eqnarray}
Thus changing the parameter $a \to a+b$ amounts to the replacement
$f \to f - g f' + O(g^2)$ with
\begin{equation}
  g(\xi) = b \, \left( \xi^3 - \frac{1}{\xi} \right) \,,
\label{g-of-xi}
\end{equation}
and such a change is generally generated by the operator $L_g$
defined by
\begin{equation}
  L_g = \oint_C \frac{d \xi}{2 \pi i} \, g(\xi) \, T(\xi) \,,
\end{equation}
where $T(\xi)$ is the energy-momentum tensor.
This can be shown by computing $[\, L_g \,, Q_f \,]$:
\begin{eqnarray}
  [\, L_g \,, Q_f \,]
  = -\oint_C \frac{d \xi}{2 \pi i} \,
  g(\xi) \, f'(\xi) \, j_B (\xi)
  + \oint_C \frac{d \xi}{2 \pi i} \left[ \,
    g(\xi) \left( \frac{f'(\xi)^2}{f(\xi)} \right)'
    +2 \, g'(\xi) \, \frac{f'(\xi)^2}{f(\xi)} \, \right] c(\xi) \,.
\end{eqnarray}
Note that
\begin{equation}
  -\frac{(f - g f')'^{\, 2}}{f - g f'}
  = -\frac{f'^{\, 2}}{f} + g \left( \frac{f'^{\, 2}}{f} \right)'
    + 2 \, g' \frac{f'^{\, 2}}{f} + O(g^2) \,.
\end{equation}
When $g(\xi)$ is given by (\ref{g-of-xi}), $L_g = b \, K_2$.
This completes the proof of (\ref{similar}).

Since the operator with $\epsilon=e^{-4 a}=1$
corresponds to the simplest operator constructed
by Takahashi and Tanimoto,
the identity (\ref{similar}) provides
an explicit relation between their operator and $Q_f (e^{-4 a})$:
\begin{equation}
  Q_f (e^{-4 a}) = e^{a K_2} \, Q_f (1) \, e^{-a K_2} \,,
\label{similarity-transformation}
\end{equation}
where
\begin{equation}
  Q_f (1) = \frac{1}{2} \, Q_B - \frac{1}{4} \, ( Q_2 + Q_{-2} )
            + 2 \, c_0 + c_2 + c_{-2} \,.
\label{Takahashi-Tanimoto-operator}
\end{equation}
Takahashi and Tanimoto constructed their kinetic operator
by expanding Witten's string field theory
around a formal exact solution.
The normalization of their solution is determined
by the condition that it solves the equation of motion
of Witten's string field theory, and it also determines
the normalization of $Q_f$ as $f(i) = 1$.
Since $Q_f (1)$ satisfies this normalization condition
and $Q_f (\epsilon)$ is related to $Q_f (1)$ by field redefinition
which preserves the cubic interaction,
the coupling constant of vacuum string field theory
should be the same as $g_T$.
However, their solution is formal because it does not have
a well-defined expression for the energy density.
The coupling constant of the theory with $Q_f (\epsilon)$
may be different from that of Witten's string field theory.
We therefore use $g$ to denote the coupling constant
of vacuum string field theory and write the action as follows:
\begin{equation}
  S = - \frac{1}{\alpha'^3 g^2} \left[
  \frac{1}{2} \langle \Psi | Q_f (\epsilon) | \Psi \rangle
  + \frac{1}{3} \langle \Psi | \Psi \ast \Psi \rangle \right] ,
\label{Q_f-action}
\end{equation}
where $Q_f (\epsilon)$ is defined by (\ref{Q_f-definition})
with $f(\xi)$ given by (\ref{f(xi)}).
It is convenient to introduce
the following dimensionless quantity ${\cal V}$
for the energy density ${\cal E}$ of the solution
representing a D25-brane:
\begin{equation}
  {\cal V} = 2 \pi^2 \alpha'^3 g^2 {\cal E} \,.
\label{V-definition}
\end{equation}
The value of ${\cal V}$ must be finite
for the string coupling constant to be finite 
because $g_T^2 = g^2 / {\cal V}$.
If $g = g_T$, we expect ${\cal V}=1$,
and if ${\cal V}$ is finite but different from 1, it indicates
$g \ne g_T$.

Finally, we comment
on the cohomology of the kinetic operator $Q_f (\epsilon)$.
According to Sen's conjecture,
there are no physical open string states at the tachyon vacuum.
The cohomology of the kinetic operator
expanded around the tachyon vacuum is
therefore predicted to be trivial.
This was confirmed numerically
for the vacuum constructed 
by level truncation \cite{Ellwood:2001py, Giusto:2003wc}.
It is a subtle question for $Q_f (\epsilon)$
with generic values of $\epsilon$
because $Q_f (\epsilon)$ maps a state in the Fock space
to a state outside the Fock space as $Q$ does.
However, $Q_f (\epsilon)$ is related
to the operator $Q_f (1)$
by a similarity transformation (\ref{similarity-transformation}),
and the cohomology of $Q_f (1)$
is well defined for states in the Fock space,
as can be seen from its explicit expression
in (\ref{Takahashi-Tanimoto-operator}).
Kishimoto and Takahashi studied the cohomology of $Q_f (1)$
in \cite{Kishimoto:2002xi},
and their conclusion was that
its cohomology is trivial in the sector with ghost number 1.\footnote
{
See also \cite{Takahashi:2003xe}.
}
This is in accord with Sen's conjecture.

\section{D25-brane solution with finite energy density}
\label{section-3}

As we discussed in the introduction and section \ref{section-2},
there must be a D25-brane solution
with finite energy density
in order for vacuum string field theory
to have a finite string coupling constant.
Let us solve the equation of motion
derived from the action (\ref{Q_f-action})
to see if there is a solution with finite energy density.
Since the kinetic operators $Q_f (\epsilon)$
for different values of $\epsilon$ are related
by field redefinition generated by $K_2$,
theories with different values of $\epsilon$ are equivalent
and should have solutions with the same energy density.

We will first solve the equation of motion by level truncation
and then solve it
in a series expansion with respect to $\epsilon$
using the method developed recently
to solve Witten's string field theory \cite{Okawa:2003zc}.\footnote
{
A different analytic framework
to solve Witten's string field theory
using the butterfly state was proposed earlier
in~\cite{Bars:2003cr}.
}
In fact, at the leading order the solution in \cite{Okawa:2003zc}
simultaneously solves the equation of motion with $Q_f (\epsilon)$
with a finite rescaling.

\subsection{Level truncation}
\label{subsection-3.1}

Level truncation is an approximation scheme
where the string field is truncated to a finite number of states
up to a certain level.
It works remarkably well in constructing the tachyon vacuum solution
in Witten's string field theory
\cite{Sen:1999nx, Moeller:2000xv, Taylor:2002fy, Gaiotto:2002wy}.
Since we are interested in a homogeneous D25-brane solution,
we only consider the zero-momentum sector.
There is only one state at level 0,
and the string field is truncated to
\begin{equation}
  | \widetilde{\Psi}^{(0)} \rangle = x \, c_1 \ket{0},
\end{equation}
where $x$ is the only parameter.
At this level, only the operators $Q_B$ and $c_0$
in $Q_f (\epsilon)$ contribute.
The coefficient in front of $Q_B$ is given
in (\ref{Q_n-coefficients}),
and the coefficient in front of $c_0$ is given by
\begin{equation}
  - \oint_C \frac{d \xi}{2 \pi i} \frac{\xi \, f'(\xi)^2}{f(\xi)}
  = \frac{1 + 3 \, \epsilon^2}{2 \, \epsilon} \,.
\end{equation}
Then the inner product
$\langle \widetilde{\Psi}^{(0)} |
Q_f (\epsilon) | \widetilde{\Psi}^{(0)} \rangle$
can be evaluated as follows:
\begin{eqnarray}
  \langle \widetilde{\Psi}^{(0)} |
  Q_f (\epsilon) | \widetilde{\Psi}^{(0)} \rangle_{density}
  &=& x^2 \,
  \langle 0 | c_{-1} \, Q_f (\epsilon) \, c_1 | 0 \rangle_{density}
\\
  &=& x^2 \, \langle 0 | c_{-1} \left(
  \frac{\epsilon}{1 + \epsilon} \, Q_B
  + \frac{1 + 3 \, \epsilon^2}{2 \, \epsilon} \, c_0
  \right) c_1 | 0 \rangle_{density}
  = \left( - \frac{\epsilon}{1 + \epsilon}
    + \frac{1 + 3 \, \epsilon^2}{2 \, \epsilon} \right) x^2.
\nonumber
\end{eqnarray}
Here and in what follows
we use the subscript $density$ to denote
a quantity divided by the volume factor of space-time.
The cubic term
$\langle \widetilde{\Psi}^{(0)}
 | \widetilde{\Psi}^{(0)} \ast \widetilde{\Psi}^{(0)} \rangle$
is the same as that of Witten's string field theory
and is given by
\begin{equation}
  \langle \widetilde{\Psi}^{(0)} | \widetilde{\Psi}^{(0)}
  \ast \widetilde{\Psi}^{(0)} \rangle_{density}
  = - \left( \frac{3 \sqrt{3}}{4} \right)^3 x^3.
\end{equation}
The dimensionless quantity ${\cal V}$
defined by (\ref{V-definition})
for the energy density ${\cal E}$
is then given by
\begin{eqnarray}
  {\cal V} &=& 2 \pi^2 \, \biggl[ \,
  \frac{1}{2} \, \langle \widetilde{\Psi}^{(0)} |
  Q_f (\epsilon) | \widetilde{\Psi}^{(0)} \rangle_{density}
  + \frac{1}{3} \, \langle \widetilde{\Psi}^{(0)}
  | \widetilde{\Psi}^{(0)}
  \ast \widetilde{\Psi}^{(0)} \rangle_{density} \, \biggr]
\nonumber \\
  &=& 2 \pi^2 \, \Biggl[ \,
  \frac{1}{2} \left( - \frac{\epsilon}{1 + \epsilon}
  + \frac{1 + 3 \, \epsilon^2}{2 \, \epsilon} \right) x^2
  - \frac{1}{3} \left( \frac{3 \sqrt{3}}{4} \right)^3 x^3 \,
  \Biggr].
\end{eqnarray}
It is a cubic function of $x$ and has a nontrivial critical point at
\begin{equation}
  x = \left( - \frac{\epsilon}{1 + \epsilon}
  + \frac{1 + 3 \, \epsilon^2}{2 \, \epsilon} \right)
  \left( \frac{4}{3 \sqrt{3}} \right)^3,
\end{equation}
and the value of ${\cal V}$ evaluated at this point is given by
\begin{equation}
  {\cal V} = \left( - \frac{\epsilon}{1 + \epsilon}
  + \frac{1 + 3 \, \epsilon^2}{2 \, \epsilon} \right)^3
  \frac{\pi^2}{3} \left( \frac{4}{3 \, \sqrt{3}} \right)^6.
\end{equation}
The energy density at the critical point
as a function of $\epsilon$
has a minimum at $\epsilon \simeq 0.6628$
with ${\cal V} \simeq 1.6843$,
and it diverges as $1/\epsilon^3$ in the limit $\epsilon \to 0$
and as $\epsilon^3$ for large $\epsilon$.
If the energy density converges to a finite value,
this level-0 solution is not a good approximation
for both small $\epsilon$ and large $\epsilon$.

At level 2, the string field is truncated to
\begin{equation}
  | \widetilde{\Psi}^{(2)} \rangle = x \, c_1 \ket{0}
  + 2 \, u \, c_{-1} \ket{0} + v \, L^m_{-2} \, c_1 \ket{0}
  - w \, b_{-2} \, c_0 \, c_1 \ket{0},
\end{equation}
where $L^m_{-2}$ is the matter part
of the Virasoro generator $L_{-2}$
and $x$, $u$, $v$, and $w$ are parameters.
In addition to $Q_B$ and $c_0$,
the operators $Q_2$, $Q_{-2}$, $c_2$, and $c_{-2}$
in $Q_f (\epsilon)$ contribute at this level.
The coefficients in front of $Q_2$ and $Q_{-2}$
are given in (\ref{Q_n-coefficients}),
and those in front of $c_2$ and $c_{-2}$
are obtained from the following integrals:
\begin{equation}
  - \oint_C \frac{d \xi}{2 \pi i} \frac{f'(\xi)^2}{\xi \, f(\xi)}
  = - \oint_C \frac{d \xi}{2 \pi i} \frac{\xi^3 \, f'(\xi)^2}{f(\xi)}
  = -\frac{1 - 3 \, \epsilon^2}{2 \, \epsilon} \,.
\end{equation}
The inner product
$\langle \widetilde{\Psi}^{(2)} |
Q_f (\epsilon) | \widetilde{\Psi}^{(2)} \rangle$
is given by
\begin{eqnarray}
  \langle \widetilde{\Psi}^{(2)} |
  Q_f (\epsilon) | \widetilde{\Psi}^{(2)} \rangle_{density}
  &=& \frac{\epsilon}{1+ \epsilon} \,
  \langle \widetilde{\Psi}^{(2)} |
  Q_B | \widetilde{\Psi}^{(2)} \rangle_{density}
  - \frac{\epsilon}{(1 + \epsilon)^2} \,
  \langle \widetilde{\Psi}^{(2)} |
  (Q_2 + Q_{-2}) | \widetilde{\Psi}^{(2)} \rangle_{density}
\nonumber \\ &&
  {}+ \frac{1 + 3 \, \epsilon^2}{2 \, \epsilon} \,
  \langle \widetilde{\Psi}^{(2)} |
  c_0 | \widetilde{\Psi}^{(2)} \rangle_{density}
  - \frac{1 - 3 \, \epsilon^2}{2 \, \epsilon} \,
  \langle \widetilde{\Psi}^{(2)} |
  (c_2 + c_{-2}) | \widetilde{\Psi}^{(2)} \rangle_{density}
\nonumber \\
  &=& \frac{\epsilon}{1+ \epsilon} \,
  (- x^2 -4 \, u^2 +13 \, v^2 +4 \, w^2 +12 \, u \, w -26 \, v \, w)
\nonumber \\ &&
  {}- \frac{2 \, \epsilon}{(1 + \epsilon)^2} \,
  x \, ( -2 \, u + 13 \, v -6 \, w )
\nonumber \\ &&
  {}+ \frac{1 + 3 \, \epsilon^2}{2 \, \epsilon} \,
  (x^2 -4 \, u^2 +13 \, v^2)
  + \frac{1 - 3 \, \epsilon^2}{\epsilon} \,
  x \, w \,.
\end{eqnarray}
The cubic term
$\langle \widetilde{\Psi}^{(2)}
 | \widetilde{\Psi}^{(2)} \ast \widetilde{\Psi}^{(2)} \rangle$
is again the same as that of Witten's string field theory
and is given by
\begin{eqnarray}
 \langle \widetilde{\Psi}^{(2)} | \widetilde{\Psi}^{(2)}
\ast \widetilde{\Psi}^{(2)} \rangle_{density}
&=& \sqrt{3} \, \Bigg(
-\frac{81}{64}x^3-\frac{99}{32}x^2\,u
-\frac{19}{16}x\,u^2-\frac{1}{8}u^3+\frac{585}{64}x^2\,v
+\frac{715}{48}x\,u\,v+\frac{1235}{432}u^2\,v~~
\nonumber\\&&\phantom{\sqrt{3} \, \Bigg(}
-\frac{7553}{192}x\,v^2-\frac{83083}{2592}u\,v^2
+\frac{272363}{5184}v^3-\frac{9}{4}x^2\,w+3\,x\,u\,w
+\frac{703}{81}u^2\,w
\nonumber\\&&\phantom{\sqrt{3} \, \Bigg(}
+\frac{65}{6}x\,v\,w
-\frac{65}{9}u\,v\,w-\frac{7553}{324}v^2\,w-x\,w^2
+\frac{94}{81}u\,w^2+\frac{65}{27}v\,w^2\Bigg) \,.
\end{eqnarray}

We now have an expression of the energy density ${\cal E}$
as a function of $x$, $u$, $v$, and $w$.
In the case of Witten's string field theory,
it is empirically known that level truncation works
very well when the Siegel gauge condition is imposed
\cite{Sen:1999nx, Moeller:2000xv, Taylor:2002fy, Gaiotto:2002wy}.
At level 2, it corresponds to setting $w=0$
and taking variations with respect to $x$, $u$, $v$.
However, it is also possible to take variations
with respect to all of $x$, $u$, $v$, and $w$
without fixing gauge.
Such an analysis was carried out
for Witten's string field theory
in \cite{Rastelli:2000iu} and \cite{Ellwood:2001ne}.
Since the Siegel gauge condition is not globally valid
\cite{Ellwood:2001ne} and we do not {\it a priori} know
if the solution can be found in the Siegel gauge
and if level truncation also works well
with the Siegel gauge condition in our case,
we carried out these two ways of level truncation
for our string field theory action.

We performed the calculation without gauge fixing up to level~2
and the calculation in the Siegel gauge up to level~6.\footnote{
We thank Wati Taylor for providing us with 
his {\it Mathematica} package.}  
In our level-6 calculation in the Siegel gauge,
the operators $Q_4$, $Q_{-4}$, $Q_6$, and $Q_{-6}$
in $Q_f (\epsilon)$ must be included
and the coefficients in front of them are given
in (\ref{Q_n-coefficients}).
On the other hand, modes of the $c$ ghost other than $c_0$
do not contribute because of the Siegel gauge condition.

\begin{figure}[ht]
\begin{center}
\epsfig{file=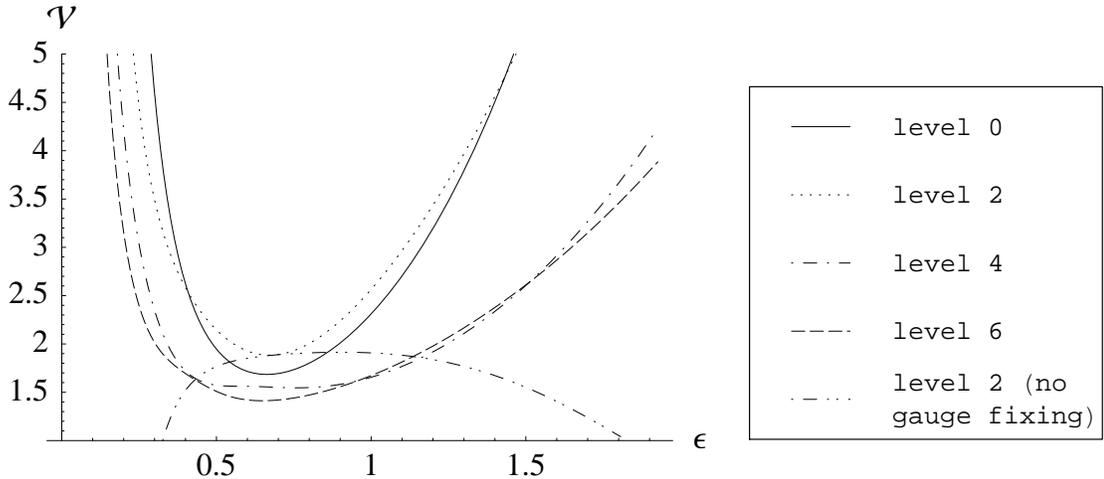}
\end{center}
\caption{Results of level truncation
in the Siegel gauge up to level~6
and level truncation without gauge fixing up to level~2.}
\label{level-6-figure}
\end{figure}

As in the case of Witten's string field theory,
we found multiple solutions for any given $\epsilon$
in all the calculations except for level~0.
When $\epsilon$ is of $O(1)$,
we found a single branch where the result is relatively 
close to that of level~0 in all cases.
The results for the dimensionless quantity ${\cal V}$
evaluated for the branch are depicted in 
figure~\ref{level-6-figure}.

While it is not clear whether the result of level truncation
converges, we can find some evidence
for the existence of a D25-brane solution
with finite energy density from the figure.
First, the result of level~2 in the Siegel gauge
is very close to that of level~2 without gauge fixing
when $\epsilon \simeq 0.65$.
This indicates that the D25-brane solution can be found
in the Siegel gauge at least when $\epsilon \simeq 0.65$.
Second, the results are reasonably stable in the range
$\epsilon \simeq 0.65$ both when we change the parameter $\epsilon$
and when we increase the level.
Not only the energy density
but also the values of the parameters such as $x$, $u$, and $v$
do not change rapidly when the level is increased.
Several values of ${\cal V}$ are tabulated
in table~\ref{level-6-truncation}. 
\begin{table}[ht]
\caption{Level truncation up to level 6.}
\label{level-6-truncation}
\begin{center}
{\renewcommand\arraystretch{1.3}
\begin{tabular}{|c||c|c|c|c|c|}
  \hline
  $\epsilon$ & Level 0 &level 2 (no gauge fixing) & level 2 
  & level 4 & level 6 \\
  \hline
  $0.2$ & $12.5016$ & $-3.11724$ & $6.39057$ & $4.22023$ & $3.15620$ \\
  \hline
  $0.4$ & $2.62057$ & $1.52154$ & $2.58262$ &  $1.72038$ & $1.69360$ \\
  \hline
  $0.6$ & $1.71580$ & $1.85186$ & $1.92678$ &  $1.56150$ & $1.42052$ \\
  \hline
  $0.8$ & $1.80139$ & $1.90758$ & $2.00495$ &  $1.54859$ & $1.47134$ \\
  \hline
  $1$ & $2.31058$ & $1.90766$ & $2.56418$ &  $1.65508$ & $1.67246$ \\
  \hline
\end{tabular}
}
\end{center}
\end{table}

Although the actions with different values of $\epsilon$
are all equivalent, level truncation does not necessarily
work equally well for all $\epsilon$.
A similar phenomenon was observed in \cite{Takahashi:2003pp},
where level truncation with different kinetic operators
obtained from $Q_B$ by field redefinition was studied.
Our results indicate that level truncation works well
when $\epsilon \simeq 0.65$.
Furthermore, the range of $\epsilon$ where ${\cal V}$ is stable
seems to become wider as the level is increased.
This qualitative behavior is in fact
what we expected from the structure of $Q_f (\epsilon)$.
When $\epsilon$ is of $O(1)$,
the coefficients in front of $Q_{2n} + Q_{-2n}$
decay fast as $n$ becomes large.
On the other hand, they do not decay fast
when $\epsilon$ is small or large.
We therefore do not expect level truncation to work well
in these regions of $\epsilon$.

As we mentioned before,
the energy density diverges as $1/\epsilon^{3}$ for small $\epsilon$
and as $\epsilon^{3}$ for large $\epsilon$ at level~0.
The numerical results in the Siegel gauge
indicate that this is also the case
at higher levels, but the coefficients in front of $\epsilon^{\pm 3}$
decrease as the level is increased.
At level~2 the coefficients are suppressed
by a factor of roughly $3.7$ compared to level~0. 
They are further suppressed by a factor of $2.5$ from level~2
to level~4 and by another factor of $2$ from level~4 to level~6.
Although we do not expect that level truncation
provides a good approximation
for small or large $\epsilon$
when the truncation level stays as low as 6
and it is not even clear if the solution exists
in the Siegel gauge for all $\epsilon$,
this tendency of suppression seems to be in favor of
the existence of a solution with finite energy density
rather than divergent energy density.

In the case of level-2 truncation without gauge fixing,
there are two other branches
where the energy density stays positive for any $\epsilon$
and diverges as $1/\epsilon^3$ for small $\epsilon$
and as $\epsilon^3$ for large $\epsilon$.
However, the smallest values of ${\cal V}$ on these branches
are approximately 36 and 110,
and the values of $x$ for these solutions are not close
to that of the level-0 solution.
On the other hand, 
the energy density ${\cal V}$ and the parameters $x$, $u$, and $v$
on the branch chosen in figure~\ref{level-6-figure}
are very close to those of the solution
in the Siegel gauge at level~2 when $\epsilon \simeq 0.65$.
An unappealing property of this branch is that 
the energy density turns negative for small $\epsilon$
and goes to zero for large $\epsilon$.
While we do not expect that level truncation
is a good approximation for these regions of $\epsilon$
when the truncation level is very low,
it would be important to confirm that these behaviors
are artifacts of truncation.

To summarize, the analysis of level truncation seems to indicate
the existence of a solution with finite energy density
$\cal V$ of $O(1)$.
This approximation scheme seems to work well
for $\epsilon \simeq 0.65$,
while it does not work well for small or large $\epsilon$,
at least when the truncation level is low.
We expect that the solution will improve
as the truncation level is increased,
and the parameter region of $\epsilon$ where level truncation
works well may become wider at higher orders.
It will not be promising, however, to study the limit
$\epsilon \to 0$ by level truncation
since higher modes of the kinetic operator are not suppressed.
We will study this region
using a different method in the next subsection.

\subsection{Solution using the butterfly state}
\label{subsection-3.2}

We have seen that level truncation does not seem to work
for small $\epsilon$.
On the other hand, it turns out that
the method developed in \cite{Okawa:2003zc}
to solve Witten's string field theory
can be used to solve the equation of motion
when $\epsilon$ is small,
providing an independent test for the existence
of a solution with finite energy density.
It is also interesting to study the case with small $\epsilon$
because this is the limit where the kinetic operator becomes
the singular midpoint $c$-ghost insertion
and we can study the contributions from subleading terms.

The ansatz for the solution
takes the form of the regulated butterfly state
with an operator insertion at the midpoint of the boundary.
Let us define $\ket{B_t ({\cal O})}$ by
\begin{equation}
  \vev{ \phi | B_t ({\cal O}) }
  = \vev{ I \circ f_t \circ \phi (0) \, {\cal O}(0) }
\end{equation}
for any state $\ket{\phi}$ in the Fock space, where
\begin{equation}
  f_t (\xi) = \frac{\xi}{\sqrt{1 + t^2 \xi^2}} \,, \qquad
  I(\xi) = -\frac{1}{\xi} \,.
\end{equation}
As in the case of the regulated butterfly state $\ket{B_t}$,
the state $\ket{B_t ({\cal O})}$ itself becomes
singular in the limit $t \to 1$,
but its inner product with a state $\ket{\phi}$ in the Fock space
has a finite limit,
which we denote by $\vev{ \phi | B ({\cal O}) }$:
\begin{equation}
  \vev{ \phi | B ({\cal O}) }
  = \lim_{t \to1 }\vev{ \phi | B_t ({\cal O}) }
  = \vev{ I \circ f_B \circ \phi (0) \, {\cal O}(0) },
\end{equation}
where
\begin{equation}
  f_B (\xi) = \frac{\xi}{\sqrt{1 + \xi^2}} \,.
\end{equation}
We will show that $\ket{B_t (c)}$
with an appropriate normalization solves
the equation of motion up to $O(\sqrt{\epsilon})$
if we take $t \to 1$ as $\epsilon \to 0$
with $\epsilon/(1-t)$ kept finite.

The quantity $\vev{ \phi | B_t (c) \ast B_t (c) }$
has been computed in \cite{Okawa:2003zc} and is given by
\begin{equation}
  \vev{ \phi | B_t (c) \ast B_t (c) }
  = \frac{4 \sqrt{2}}{3^{1/4}} \sqrt{1-t}
    \vev{ \phi | B(c \partial c) }
    + O \left( (1-t)^{3/2} \right).
\label{phi-B_t(c)-B_t(c)}
\end{equation}
Let us compute $\vev{ \phi | Q_f (\epsilon) | B_t (c) }$.
The function $f(\xi)$ can be written in the coordinate
$z = I \circ f_t (\xi)$ as
\begin{equation}
  f(z) = f(\xi(z))
  = -\frac{\epsilon^2 [z^2-(1+t^2)]^2}
          {[(1+\epsilon)(z^2-t^2)+(1-\epsilon)]
           [(1-\epsilon)(z^2-t^2)+(1+\epsilon)]} \,.
\end{equation}
The double zeros are at $z = \pm \sqrt{1+ t^2}$,
the poles are at
\begin{equation}
  z^2 = t^2 - \frac{1-\epsilon}{1+\epsilon} \,, \qquad
  z = \pm i \sqrt{\frac{1+\epsilon}{1-\epsilon}-t^2} \,,
\end{equation}
and the open-string midpoint is at $z = \pm i \sqrt{1-t^2}$.
The action of the kinetic operator $Q_f (\epsilon)$
on an operator at the origin
${\cal O } (0)$ is given by
\begin{equation}
  Q_f (\epsilon) \cdot {\cal O } (0) \equiv
  \oint \frac{dz}{2 \pi i} \biggl[ \, f(z) \, j_B (z) \, {\cal O } (0)
  -  \frac{f'(z)^2}{f(z)} \, c(z) \, {\cal O } (0) \, \biggr] \,,
\end{equation}
where the contour encircles
the origin and the poles
at $z^2 = t^2 - (1-\epsilon)/(1+\epsilon)$ counterclockwise.
The other poles are located outside the contour.
Let us define
\begin{equation}
  p^2 = t^2 - \frac{1-\epsilon}{1+\epsilon} \,.
\label{poles-in-z}
\end{equation}
Then $f(z)$ can be written
in the same form as (\ref{f-of-xi}) using $p$:
\begin{equation}
  f(z) = - \frac{\epsilon^2}{1-\epsilon^2}
  \frac{\left( z^2 - p^2
  - \displaystyle{\frac{2}{1+\epsilon}} \right)^2}
  { (z^2-p^2)
  \left( z^2 - p^2
  + \displaystyle{\frac{\strut 4 \, \epsilon}{1-\epsilon^2}}
  \right)} \,.
\label{f-of-z}
\end{equation}
Note that $f(z)$ depends on $t$ only through $p$.
The poles inside the contour are located at $\pm p$.
The value of $p^2$ can be both positive and negative
depending on $t$ and $\epsilon$.
In particular, when $p^2=0$, the two single poles become
a double pole at the origin. We have to study this case separately.

Let us first consider the case where $p^2 \ne 0$.
Since $f(\xi)$ in (\ref{f-of-xi}) and $f(z)$ in (\ref{f-of-z})
take the same form, $f(z)$ and $-f'(z)^2/f(z)$ can be expanded
about $z = \pm p$
as in (\ref{f-of-xi-expansion}) and (\ref{f'^2/f-of-xi-expansion}),
respectively, with $\xi$ replaced by $z$.
Consequently, $Q_f (\epsilon)$ in the $z$ coordinate
can be written in the same form as (\ref{Q_f-expansion})
with $\xi$ replaced by $z$.
It is given by local insertions of $j_B(\pm p)$, $c(\pm p)$, 
$\partial c(\pm p)$ and $\partial^2 c(\pm p)$
with $p$ defined by (\ref{poles-in-z})
and integrals around the origin.
When $\epsilon$ is small and $t$ is close to 1, $p^2$ is small.
In this case, the action of $Q_f (\epsilon)$
on an operator at the origin can be evaluated using OPE's.
It is convenient to introduce a parameter $\eta$ defined by
\begin{equation}
  \eta = \frac{1 + t^2}{1-t^2} \, \epsilon \,,
\label{eta}
\end{equation}
and we will consider the case where $\eta$ is of $O(1)$.\footnote
{
The following method is applicable as long as $p^2$ is small
and does not necessarily require $\eta$ to be of $O(1)$.
However, we have to make sure that
subleading terms in $\epsilon$ are suppressed.
Furthermore, $\eta$ is constrained by the requirement
that the equation of motion contracted
with the solution itself be solved with good accuracy,
as we will discuss later.
We found a meaningful solution
only when $\eta$ is of $O(1)$.
We will also present a picture at the end of this subsection
which naturally explains why we should consider
the case with $\eta$ of $O(1)$.
}
Namely, $\epsilon$ and $1-t$ are of the same order
and related by
\begin{equation}
  \epsilon = \eta \, (1-t) + O( (1-t)^2 ) \,, \qquad
  1-t = \frac{\epsilon}{\eta} + O(\epsilon^2) \,.
\end{equation}
In this case, $p^2$ is given by
\begin{equation}
  p^2 = \frac{2 \, \epsilon \, (\eta -1)}
             {(\eta + \epsilon)(1 + \epsilon)}
  = 2 \left( 1-\frac{1}{\eta} \right) \epsilon + O (\epsilon^2) \,.
\end{equation}
Therefore, $p$ is of $O(\sqrt{\epsilon})$.
In the computation of
$\langle \phi | Q_f (\epsilon) | B_t (c) \rangle$,
the operator inserted at the origin is $c$,
and the relevant OPE's are
\begin{eqnarray}
  j_B (z) \,  c (0)
  &=& \frac{1}{z} \, c \partial c (0)
  +O(z^0)\,,
\\
c (z) \, c (0)
  &=&  - z \, c \partial c (0)
  +O(z^2)\,,
\\
\partial c (z) \, c (0)
  &=&  - c \partial c (0)
  +O(z)\,,
\\
\partial^2 c (z) \, c (0)
  &=&  - c \partial^2 c (0)
  +O(z)\,.
\end{eqnarray}
We compute the leading term
of $\langle \phi | Q_f (\epsilon) | B_t (c) \rangle$
for small $\epsilon$, which is of $O(\epsilon^0)$.
The relevant part of $Q_f (\epsilon) \cdot c(0)$
to this order is
\begin{eqnarray}
  Q_f (\epsilon) \cdot c(0) &=& - \frac{\epsilon}{2 \, p }
             ( j_B (p) - j_B (-p) ) \, c(0)
        + \oint_{C_0} \frac{d z}{2 \pi i} \, f(z) \, j_B (z) \, c(0)
  + \frac{p^2 + \epsilon }{4 \, p^2 }
        ( \partial c (p) + \partial c(-p) ) \, c(0)
\nonumber \\ &&
  - \frac{p^4 - \epsilon \, p^2 + \epsilon^2}
             {4 \, \epsilon \, p^3}
        ( c (p) - c(-p) ) \, c(0)
  - \oint_{C_0} \frac{d z}{2 \pi i} \,
  \frac{f'(z)^2}{f(z)} \, c(z) \, c(0) + O(\epsilon) \,,
\end{eqnarray}
where the contour $C_0$ encircles the origin counterclockwise.
The functions $f(z)$ and $-f'(z)^2/f(z)$ are expanded
about the origin when $p^2 \ne 0$ as follows:
\begin{eqnarray}
  f(z) &=& - \frac{\epsilon^2 \, (2 + p^2 + \epsilon \, p^2)^2}
  {p^2 \, (1 + \epsilon)^2 \,
  (p^2 - 4 \, \epsilon - \epsilon^2 \, p^2)} + O(z^2)
  = \left[ \, - \frac{4 \, \epsilon^2}{p^2 \, (p^2 - 4 \, \epsilon)}
  +O(\epsilon) \, \right] + O(z^2) \,,
\label{f(z)-expansion}
\\
  - \frac{f'(z)^2}{f(z)} &=& O(z^2) \,.
\end{eqnarray}
Therefore, $Q_f (\epsilon) \cdot c(0)$ is given by
\begin{eqnarray}
  Q_f (\epsilon) \cdot c(0)
  &=& \left[ \, - \frac{\epsilon}{p^2}
  - \frac{4 \, \epsilon^2}{p^2 \, (p^2 - 4 \, \epsilon)}
  - \frac{p^2 + \epsilon}{2 \, p^2 }
  + \frac{p^4 - \epsilon \, p^2 + \epsilon^2}
             {2 \, \epsilon \, p^2} \, \right] \,
  c \partial c (0) + O(\epsilon)
\nonumber \\
  &=& \left[ \, \frac{\eta}{2 \, (1 + \eta)}
  - \frac{1}{\eta} \, \right]
  c \partial c (0) + O(\epsilon) \,,
\end{eqnarray}
and $\langle \phi | Q_f (\epsilon) | B_t (c) \rangle$ is
\begin{equation}
  \langle \phi | Q_f (\epsilon) | B_t (c) \rangle
  = \left[ \, \frac{\eta}{2 \, (1 + \eta)}
  - \frac{1}{\eta} \, \right]
  \langle \phi | B_t (c \partial c) \rangle + O(\epsilon)
  = \left[ \, \frac{\eta}{2 \, (1 + \eta)}
  - \frac{1}{\eta} \, \right]
  \langle \phi | B (c \partial c) \rangle + O(\epsilon) \,.
\label{phi-Q_f-B_t(c)}
\end{equation}
While the operator $Q_f (\epsilon)$
in $\langle \phi_1 | Q_f (\epsilon) | \phi_2 \rangle$
with $| \phi_1 \rangle$ and $| \phi_2 \rangle$ in the Fock space
is dominated by the midpoint $c$-ghost insertion
when $\epsilon$ is small,
this is not the case
for $\langle \phi | Q_f (\epsilon) | B_t (c) \rangle$.
Contributions from subleading terms in $Q_f (\epsilon)$
are nonvanishing in this computation.

Let us next consider the case where $p^2 = 0$,
which corresponds to $\eta = 1$.
In this case, $t$ and $\epsilon$ are related by
\begin{equation}
  t = \sqrt{\frac{1-\epsilon}{1+\epsilon}} \,,
\label{t-epsilon}
\end{equation}
and the two single poles inside the contour become a double pole
at the origin.
The computation in fact simplifies in this case.
The functions $f(z)$ and $-f'(z)^2/f(z)$ are expanded
about the origin as follows:
\begin{eqnarray}
  f(z) &=& - \frac{\epsilon}{(1+\epsilon)^2} \frac{1}{z^2}
    + \frac{1 + 3 \, \epsilon}{4 \, (1 + \epsilon)}
    - \frac{(1 + \epsilon)^2}{16 \, \epsilon} \, z^2
    + O(z^4) \,,
\\
  - \frac{f'(z)^2}{f(z)}
  &=& \frac{4 \, \epsilon}{(1+\epsilon)^2} \frac{1}{z^4}
    + \frac{1 + 3 \, \epsilon}{1 + \epsilon} \frac{1}{z^2}
    + \left( - \frac{1}{2 \, \epsilon}
    + \frac{3 \, \epsilon}{2} \right) + O(z^2) \,.
\end{eqnarray}
Therefore, $Q_f (\epsilon) \cdot c(0)$ is
\begin{equation}
  Q_f (\epsilon) \cdot c(0)
  = - \frac{3 \, (1 + 3 \, \epsilon)}{4 \, (1 + \epsilon)} \,
      c \partial c (0) + O(\epsilon)
  = - \frac{3}{4} \, c \partial c (0) + O(\epsilon) \,,
\end{equation}
and $\vev{ \phi | Q_f (\epsilon) | B_t (c) }$ is
\begin{equation}
  \vev{ \phi | Q_f (\epsilon) | B_t (c) }
  = - \frac{3}{4} \vev{ \phi | B_t (c \partial c) } + O(\epsilon)
  = - \frac{3}{4} \vev{ \phi | B (c \partial c) } + O(\epsilon) \,.
\end{equation}
This coincides with (\ref{phi-Q_f-B_t(c)}) when $\eta =1$.
Therefore, (\ref{phi-Q_f-B_t(c)}) is valid for any $\eta$
including $\eta =1$.

The leading term of
$\langle \phi | Q_f (\epsilon) | B_t (c) \rangle$
is proportional to that of $\vev{ \phi | B_t (c) \ast B_t (c) }$
in (\ref{phi-B_t(c)-B_t(c)}). If we define
\begin{equation}
  | \Psi^{(0)} \rangle = x \ket{ B_t (c) }
\label{normalize-Psi0}
\end{equation}
with
\begin{equation}
 x = -\frac{3^{1/4} \sqrt{\eta}}{4 \sqrt{2}}
  \left[ \frac{\eta}{2 \, (1 + \eta)} - \frac{1}{\eta} \right]
  \frac{1}{\sqrt{\epsilon}}
\qquad\hbox{and}\qquad
 t = \sqrt{\frac{\eta-\epsilon}{\eta+\epsilon}} \,,
\label{define-x}
\end{equation}
the state $| \Psi^{(0)} \rangle$ solves the equation of motion
up to $O(\sqrt{\epsilon})$:
\begin{equation}
  \langle \phi | Q_f (\epsilon) | \Psi^{(0)} \rangle
  + \langle \phi | \Psi^{(0)} \ast \Psi^{(0)} \rangle
  = O(\sqrt{\epsilon})
\end{equation}
for any state $\ket{\phi}$ in the Fock space.
If we take the limit $\epsilon \to 0$,
the state $| \Psi^{(0)} \rangle$
formally solves the equation of motion exactly.
However, the state $\ket{B_t (c)}$ becomes
singular and the coefficient diverges as $1/\sqrt{\epsilon}$
in the limit so that
we do not intend to take the strict $\epsilon \to 0$ limit.
On the other hand, the state $| \Psi^{(0)} \rangle$
is well defined as long as $\epsilon$ is finite.
If we choose $\epsilon$ to be 0.0001, for example,
the equation of motion is solved with fairly good precision
for any state $\ket{\phi}$ in the Fock space.

Since the solution is outside the Fock space,
it is a nontrivial question if the equation of motion
can be satisfied when it is contracted with the solution itself.
We can study this by evaluating the following dimensionless
quantity:
\begin{equation}
  {\cal R} [\Psi] \equiv
  - \frac{\langle \Psi | Q_f (\epsilon) | \Psi \rangle}
         {\langle \Psi | \Psi \ast \Psi \rangle} \,.
\end{equation}
If ${\cal R}$ for the solution is close to $1$,
the equation of motion contracted the solution itself
is satisfied with good accuracy.
It is also a nontrivial question
if the normalized energy density ${\cal V}$
is well defined and finite.
Let us compute ${\cal R}$ and ${\cal V}$
for the solution $| \Psi^{(0)} \rangle$:
\begin{eqnarray}
  {\cal R} [ \Psi^{(0)} ]
  &=& - \frac{\langle B_t (c) | Q_f (\epsilon) | B_t (c) \rangle}
        {x \, \langle B_t (c) | B_t (c) \ast B_t (c) \rangle} \,,
\\
  {\cal V} [ \Psi^{(0)} ]
  &=& 2 \pi^2 \left[
  \frac{x^2}{2} \langle B_t (c) | Q_f (\epsilon) | B_t (c)
  \rangle_{density}
  + \frac{x^3}{3} \langle B_t (c) | B_t (c) \ast B_t (c)
  \rangle_{density} \right] \,,
\end{eqnarray}
with $x$ and $t$ given in (\ref{define-x}).
The quantity $\vev{ B_t (c) | B_t (c) \ast B_t (c) }$ has been
computed in \cite{Okawa:2003zc} and is given by
\begin{equation}
  \vev{ B_t (c) | B_t (c) \ast B_t (c) }_{density}
  = - \left( \frac{3 \sqrt{3}}{4} \right)^3
        (1-t^4)^{3/2}.
\label{cubic-Psi0}
\end{equation}
We need to compute the inner product
$\vev{ B_t (c) | Q_f (\epsilon) | B_t (c) }$.
The computation is closely related to that of
$\vev{ B_t (c) | Q_B | B_t (c) }$ in \cite{Okawa:2003zc}.

When we deal with the star multiplication
of the regulated butterfly state,
it is convenient to use the $\hat{z}$ coordinate
\cite{Gaiotto:2002kf} defined by
\begin{equation}
  \hat{z} = \arctan \xi \,.
\end{equation}
In the $\hat{z}$ coordinate,
the left and right halves of the open string
of the regulated butterfly state
are mapped to semi-infinite lines
parallel to the imaginary axis in the upper-half plane.
Thus gluing can be performed simply by translation.
We glue two regulated butterfly states together in this way.
We then map the resulting surface to an upper-half plane
by a conformal transformation,
and the coordinate of the upper-half plane was called $z$
in \cite{Okawa:2003cm}.\footnote{
The $z$ coordinate here should not be confused
with the coordinate $z = I \circ f_t (\xi)$ we used before.}
The relation between the $z$ coordinate and the $\xi$ coordinate
can be read off from (C.15) and (C.16) in \cite{Okawa:2003cm}
and is given by
\begin{equation}
  \left( \frac{2 \, \xi}{1- \xi^2} \right)^2
  = \frac{(1-z^2)^2-4 \, q^2 \, z^2}{4 \, (1+q^2) \, z^2} \,,
\label{xi-z}
\end{equation}
where
\begin{equation}
  q = \frac{2 \, t}{1-t^2} \,.
\end{equation}
Since $f(\xi)$ can be written as
\begin{equation}
  f(\xi) = \left[ 1 - \frac{1}{\epsilon^2} \left\{
                  1 + \left( \frac{2 \, \xi}{1- \xi^2} \right)^2
           \right\} \right]^{-1},
\end{equation}
the form of $Q_f (\epsilon)$ in the $z$ coordinate can be easily
derived using (\ref{xi-z}).
The expression of the inner product
$\vev{ B_t (c) | Q_f (\epsilon) | B_t (c) }$
further simplifies in the $w$ coordinate defined by
\begin{equation}
  w = \frac{z-1}{z+1} \,.
\label{w-coordinate}
\end{equation}
The function $f(\xi)$ can be written in the $w$ coordinate
as follows:
\begin{eqnarray}
  f(w) &=& f(\xi(w))
  = \left[ 1 - \frac{(1 - t^2)^2}{\epsilon^2 \, (1 + t^2)^2}
    \left\{ 1 + \left( \frac{2 \, w}{1- w^2} \right)^2 \right\}
    \right]^{-1}
\nonumber \\
  &=& \left[ 1 - \frac{1}{\eta^2}
      \left\{ 1 + \left( \frac{2 \, w}{1- w^2} \right)^2 \right\}
      \right]^{-1},
\end{eqnarray}
where we have used the parameter $\eta$ introduced in (\ref{eta}).
The conformal factor of the $c$ ghost
associated with the mapping to the $z$ coordinate
has been computed in \cite{Okawa:2003cm, Okawa:2003zc}.
The two $c$ ghosts are mapped to
\begin{equation}
  \frac{\sqrt{1-t^4}}{2} \, c(\pm 1)
\end{equation}
in the $z$ coordinate. They are further mapped to
$\sqrt{1-t^4} \, c(0)$ and $\sqrt{1-t^4} \, I \circ c(0)$
in the $w$ coordinate.
Therefore, the computation of
$\vev{ B_t (c) | Q_f (\epsilon) | B_t (c) }$
reduces to that of $\vev{ 0 | c_{-1} Q_f (\eta) c_1 | 0 }$.
The latter has already been calculated in the previous subsection
so that $\vev{ B_t (c) | Q_f (\epsilon) | B_t (c) }$ is given by
\begin{eqnarray}
  \vev{ B_t (c) | Q_f (\epsilon) | B_t (c) }_{density}
  &=& (1 - t^4)
  \langle 0 | c_{-1} Q_f (\eta) c_1 | 0 \rangle_{density}
\nonumber \\
  &=& (1 - t^4)
    \left( - \frac{\eta}{1 + \eta}
    + \frac{1 + 3 \, \eta^2}{2 \, \eta} \right) .
\label{kinetic-Psi0}
\end{eqnarray}

We are now ready to evaluate ${\cal R} [\Psi^{(0)}]$
and ${\cal V} [\Psi^{(0)}]$.
Combining (\ref{define-x}), (\ref{cubic-Psi0}),
and (\ref{kinetic-Psi0}), they are given by
\begin{eqnarray}
  {\cal R} [\Psi^{(0)}]
  &=& 2^{15/2} \cdot 3^{-19/4}
  \left( \frac{1+3 \eta^2}{2 \eta} -\frac{\eta}{1+\eta} \right)
  \left( \frac{1}{\eta} - \frac{\eta}{2 (1+\eta)} \right)^{-1}
  + O(\epsilon) \,,
\\
  {\cal V} [\Psi^{(0)}]
  &=& 2 \pi^2
  \left( \frac{1}{\eta} - \frac{\eta}{2 (1+\eta)} \right)^{2}
  \left[ \frac{\sqrt{3}}{16}
  \left( \frac{1+3 \eta^2}{2 \eta} -\frac{\eta}{1+\eta} \right)
  - 2^{-21/2} \cdot 3^{17/4}
  \left( \frac{1}{\eta} - \frac{\eta}{2 (1+\eta)} \right)
  \right] + O(\epsilon) \,.
\nonumber \\
\end{eqnarray}
For finite $\eta$ the ratio ${\cal R} [\Psi^{(0)}]$ is of $O(1)$
and has the value 
${\cal R} [\Psi^{(0)}] = 1$ when $\eta \simeq 0.657104$.
This means that while the equation of motion
contracted with a state in the Fock space can be solved
up to $O(\sqrt{\epsilon})$ for any $\eta$,
the equation of motion contracted with the solution itself
can be satisfied only when $\eta \simeq 0.657104$.
We therefore think that our ansatz works well
only when $\eta \simeq 0.657104$.
The energy density ${\cal V}$
is well defined and finite even in the limit $\epsilon \to 0$.
Its value in the limit $\epsilon \to 0$
when $\eta \simeq 0.657104$ is ${\cal V} \simeq 1.68454$.
This is close to the results obtained by level truncation
for finite $\epsilon$,
while we are now looking at $Q_f (\epsilon)$ with small $\epsilon$.
This is exactly what we expected
since the actions with different values of $\epsilon$
are equivalent and the energy density should be
independent of $\epsilon$.
It is also interesting to note that
${\cal V} [\Psi^{(0)}]$ in the limit $\epsilon \to 0$
as a function of $\eta$ has a stationary point
at $\eta \simeq 0.653151$
where ${\cal V} [\Psi^{(0)}] \simeq 1.6847$.
This is very close to the point where ${\cal R} [\Psi^{(0)}] =1$.
The energy density is therefore rather stable
in the region of $\eta$ where our ansatz is valid.

We have also computed the solution at the next-to-leading order.
The computations are summarized in appendix B.
As in the case of solving Witten's string field theory
using this method \cite{Okawa:2003zc},
we found many solutions for a given $\eta$.
While the dimensionless ratio ${\cal R}$ is generically
not close to $1$, we found one branch of solutions
where ${\cal R} \simeq 1$.
The result is summarized in table \ref{ratio-tension-eta}.
Remarkably, ${\cal R}$ becomes close to $1$
around $\eta \simeq 0.65$
as in the case of the leading-order solution.
The energy density at $\eta \simeq 0.65$
is ${\cal V} \simeq 3.14$ and seems a little too large
compared with the leading-order result
or with the results from level truncation.
However, it is in fact quite nontrivial
to have obtained the result of $O(1)$ in this numerical analysis
since solutions on other branches are generically
far away from $O(1)$.

\begin{table}[ht]
\caption{${\cal R}$ and ${\cal V}$
for the solution at the next-to-leading order.}
\label{ratio-tension-eta}
\begin{center}
{\renewcommand\arraystretch{1.1}
\begin{tabular}{|c||c|c|}
\hline
$\eta$ & ${\cal R}$ & ${\cal V}$ \\
\hline
$0.55$ & $0.889736$ & $2.89180$ \\
\hline
$0.60$ & $0.938201$ & $3.02575$ \\
\hline
$0.65$ & $1.00333$ & $3.14301$ \\
\hline
$0.70$ & $1.09325$ & $3.21288$ \\
\hline
$0.75$ & $1.22089$ & $3.20464$ \\
\hline
$0.80$ & $1.41254$ & $3.08059$ \\
\hline
$0.85$ & $1.73586$ & $2.77125$ \\
\hline
$0.90$ & $2.44342$ & $2.10526$ \\
\hline
\end{tabular}
}
\end{center}
\end{table}

As was discussed in \cite{Okawa:2003cm},
the finiteness of the energy density of a solution
depends on the subleading structure of the kinetic operator
(\ref{Q-conjecture}).
We have explicitly demonstrated that
our kinetic operator does admit
a perturbative solution with finite energy density.
Level truncation
and the perturbative method using the butterfly state
are valid in different parameter regions of $\epsilon$,
but both gave qualitatively compatible results
with ${\cal V}$ of $O(1)$.
We regard this as evidence
for the existence of a D25-brane solution
with finite energy density.

It was also argued in \cite{Okawa:2003cm} that
contributions to $\vev{ \Psi | Q_f (\epsilon) | \Psi }$ from
subleading terms of the kinetic operator
can be of the same order as that from the leading term.
It is indeed the case for our kinetic operator.
Let us replace $Q_f (\epsilon)$
in $\vev{ B_t (c) | Q_f (\epsilon) | B_t (c) }$
by its leading term in (\ref{Q-singular-limit}).
Since $\vev{ B_t (c) | Q | B_t (c) }_{density} = (1-t^2)^2$
as computed in \cite{Okawa:2003zc},
the contribution to $\vev{ B_t (c) | Q_f (\epsilon) | B_t (c) }$ from
the leading term of $Q_f (\epsilon)$ is given by
\begin{equation}
  \frac{1}{2 \, \epsilon} \vev{ B_t (c) | Q | B_t (c) }_{density}
  = \frac{1-t^4}{2 \, \eta} \,.
\end{equation}
This is different from (\ref{kinetic-Psi0}).
Because subleading terms of $Q_f (\epsilon)$ contribute
to $\vev{ \phi | Q_f (\epsilon) | \Psi }$
and $\vev{ \Psi | Q_f (\epsilon) | \Psi }$,
the ratio (\ref{compatibility}) depends on $\eta$
and can be close to 1.
We have seen that
the ratio (\ref{compatibility}) is in fact close to 1
when $\eta \simeq 0.65$, and thus the problem of
the incompatibility between (\ref{equation-contracted-with-Psi})
and (\ref{equation-contracted-with-phi}) has been resolved
by subleading terms of $Q_f (\epsilon)$.

Level truncation works well
when $\epsilon \simeq 0.65$
and the solution based on the butterfly state works well
when $\eta \simeq 0.65$.
It it interesting to note that these two numbers are so close.
This might be explained by the following picture.
Let us denote the exact solution in the Siegel gauge
at $\epsilon_\ast = 0.65$ by $\ket{\Psi_\ast}$.
Our results of level truncation
seem to indicate that $\ket{\Psi_\ast}$
is relatively close to a state in the Fock space.
Since the kinetic operators with different values of $\epsilon$
are related by the similarity transformation (\ref{similar}),
exact solutions for different values of $\epsilon$
are given by $e^{\, b K_2} \ket{\Psi_\ast}$, where
\begin{equation}
  b = \frac{1}{4} \ln \frac{\epsilon_\ast}{\epsilon} \,.
\end{equation}
The action of the operator $e^{\, b K_2}$ on $\ket{\Psi_\ast}$
would be nontrivial, especially for small $b$.
However, the state $e^{\, b K_2} \ket{\Psi_\ast}$
with large $b$ might be approximated in the following way.
It was shown in \cite{Schnabl:2002ff} that the action of 
$e^{\, b K_2}$ on the $SL(2,R)$-invariant vacuum $\ket{0}$
gives the regulated butterfly state $\ket{B_t}$.
The relation between the parameters $b$ and $t$
can be read off from
\begin{equation}
  e^{\, b K_2} \ket{0}
  = \exp \left[ -\frac{\tanh (2 b)}{2} L_{-2} \right] \ket{0}
\label{K2-butterfly}
\end{equation}
and
\begin{equation}
  \ket{B_t} = \exp \left( -\frac{t^2}{2} L_{-2} \right) \ket{0} \,.
\label{L-2-butterfly}
\end{equation}
The operator $e^{\, b K_2}$ with large $b$ therefore
transforms the vacuum $\ket{0}$ into a regulated butterfly state
$\ket{B_t}$ with $t$ close to 1.
Since $\ket{\Psi_\ast}$ is expected to be close
to a state in the Fock space,
the state $e^{\, b K_2} \ket{\Psi_\ast}$ with large $b$
might be approximated
by a regulated butterfly state $e^{\, b K_2} \ket{0}$
with operator insertions.
 From (\ref{K2-butterfly}) 
and (\ref{L-2-butterfly}) we can 
estimate the parameter $t$ of the butterfly state to be
\begin{equation}
  t^2 = \tanh \left( \frac{1}{2}
  \ln \frac{\epsilon_\ast}{\epsilon} \right)
  = \frac{\epsilon_\ast - \epsilon}{\epsilon_\ast + \epsilon} \,.
\end{equation}
Then, $\eta$ defined in (\ref{eta}) precisely coincides
with $\epsilon_\ast$:
\begin{equation}
  \eta = \frac{1 + t^2}{1-t^2} \, \epsilon = \epsilon_\ast \,.
\end{equation}
This argument is too simplified,
but it could be a qualitative explanation
of the coincidence of the two numbers.
This picture also naturally explains why we should consider
the case with $\eta$ of $O(1)$.

The parameter $\epsilon$ of $Q_f (\epsilon)$
and the parameter $t$ of $\ket{B_t}$ are both changed
by a transformation generated by $K_2$.
The butterfly state is special for $Q_f (\epsilon)$
among other star-algebra projectors in this respect.
The method of \cite{Okawa:2003zc}
based on the butterfly state
has been generalized in \cite{Yang:2004xz} to use other projectors,
and such generalization would be useful
for other choices of the function $f$ in $Q_f$.

\section{Construction of other D-brane solutions}
\label{section-4}

We have obtained evidence that the string field theory
with $Q_f (\epsilon)$ has a D25-brane solution
with finite energy density.
Let us now consider other D-brane solutions.
In \cite{Rastelli:2001vb}, other D-brane solutions
were constructed by changing the boundary condition
of the surface state.
Since the equation of motion no longer factorizes
into the matter and ghost sectors in our case, 
it may seem that the construction of other D-brane solutions
becomes nontrivial.
We show that changing
the boundary condition of the surface state
works in our case as well
because of a special property of the function $f(\xi)$.

Let us consider our solution based on the regulated butterfly state.
Following \cite{Rastelli:2001vb}, we insert
the following pair of boundary condition changing operators
in the $\hat{z}$ coordinate:
\begin{equation}
  \sigma^+ \left( \frac{\pi}{4} + \varepsilon \right) \,, \qquad
  \sigma^- \left( -\frac{\pi}{4} - \varepsilon \right) \,,
\end{equation}
where $\varepsilon$ is a regularization parameter to be sent to 0.
We also need to multiply the state
by a factor $(2 \varepsilon)^{2h}$ to solve the equation of motion,
where $h$ is the conformal dimension of $\sigma^{\pm}$.
After mapping the operator to the coordinate $I \circ f_t (\xi)$
and an appropriate rescaling of $\varepsilon$,
our ansatz for the solution can be written as
\begin{equation}
  \vev{ \phi | B_t^\sigma ({\cal O}) }
  = (2 \varepsilon)^{2h}
  \vev{ I \circ f_t \circ \phi (0) \, 
  \sigma^+ \left( -\sqrt{1+t^2} + \varepsilon \right) \,
  {\cal O}(0) \,
  \sigma^- \left( \sqrt{1+t^2} - \varepsilon \right) }
\end{equation}
for any state $\ket{\phi}$ in the Fock space.
Let us consider
$\langle \phi | Q_f (\epsilon) | B_t^\sigma ({\cal O}) \rangle$
and $\langle \phi
| B_t^\sigma ({\cal O}) \ast B_t^\sigma ({\cal O}) \rangle$.
A pair of $\sigma^+$ and $\sigma^-$ approach to each other
in $\langle \phi
| B_t^\sigma ({\cal O}) \ast B_t^\sigma ({\cal O}) \rangle$,
and the leading term in their OPE 
cancels one of the two factors of $(2 \varepsilon)^{2h}$.
The remaining part of the computation is exactly the same
except that the boundary condition has been changed.
In the computation of
$\langle \phi | Q_f (\epsilon) | B_t^\sigma ({\cal O}) \rangle$,
the action of $Q_f (\epsilon)$ on $\sigma^{\pm}$ can potentially
produce nonvanishing contributions.
Since both $f(z)$ and $f'(z)^2/f(z)$ are regular
at the insertion points of $\sigma^{\pm}$
and the OPE between $c$ and $\sigma^{\pm}$ is trivial,
$Q_f (\epsilon) \cdot \sigma^{\pm}$ is given by
\begin{eqnarray}
  Q_f (\epsilon) \cdot \sigma^{\pm} (w)
  &=& \oint_{C_w} \frac{dz}{2 \pi i} \,
  f(z) \, j_B (z) \, \sigma^{\pm} (w)
\nonumber \\
  &=& \oint_{C_w} \frac{dz}{2 \pi i} \, f(z) \left\{
  \frac{h}{(z-w)^2} \, c \sigma^{\pm} (w)
  + \frac{1}{z-w} \left[ h (\partial c) \sigma^{\pm} (w)
  + c \partial \sigma^{\pm} (w) \right] \right\}
\nonumber \\
  &=& h \, f'(w) \, c \sigma^{\pm} (w)
  + f(w) \left[ h (\partial c) \sigma^{\pm} (w)
  + c \partial \sigma^{\pm} (w) \right] \,,
\end{eqnarray}
where the contour $C_w$ encircles $w$ counterclockwise.
Since $f(z)$ has a double pole at both ends of the open string,
$Q_f (\epsilon) \cdot \sigma^{\pm}$ vanishes
in the limit $\varepsilon \to 0$.
More explicitly, from the expression of $f(z)$
in the coordinate $z = I \circ f_t (\xi)$ we find
\begin{eqnarray}
  f(z) \Bigr|_{z= \mp \sqrt{1+t^2} \pm \varepsilon}
  = -(1+t^2) \epsilon^2 \varepsilon^2 + O(\varepsilon^3) \,,
\nonumber \\
  f'(z) \Bigr|_{z= \mp \sqrt{1+t^2} \pm \varepsilon}
  = \mp 2(1+t^2) \epsilon^2 \varepsilon + O(\varepsilon^2) \,.
\end{eqnarray}
Therefore, if $| B_t ({\cal O}) \rangle$ solves the equation
\begin{equation}
  \langle \phi | Q_f (\epsilon) | B_t ({\cal O}) \rangle
  + \langle \phi | B_t ({\cal O}) \ast B_t ({\cal O}) \rangle = 0
\end{equation}
for any $\ket{\phi}$ in the Fock space,
$| B_t^\sigma ({\cal O}) \rangle$ solves the same equation
in the limit $\varepsilon \to 0$:
\begin{equation}
  \langle \phi | Q_f (\epsilon) | B_t^\sigma ({\cal O}) \rangle
  + \langle \phi | B_t^\sigma ({\cal O})
  \ast B_t^\sigma ({\cal O}) \rangle = 0 \,.
\end{equation}
As in the case of \cite{Rastelli:2001vb},
we interpret solutions generated in this way
as D-branes with the boundary condition
associated with $\sigma^{\pm}$.

This mechanism of generating new solutions in fact works
for any solution which consists of a surface state
with insertions of $b$, $c$,
and the matter part of the energy-momentum tensor
which we denote by $T^m$.
Moreover, it works for any kinetic operator
which is made of $T^m$, $b$, and $c$ and whose action
on the boundary condition changing operators vanishes.
In particular, it works for the level-truncation solution
for $Q_f (\epsilon)$ with $\epsilon$ of $O(1)$.
On the other hand, this mechanism does not work for
the BRST operator $Q_B$.
It corresponds to the case where $f(\xi)$ is constant
and the action of $Q_B$
on the boundary condition changing operators does not vanish.
We are also not able to make use of this method
to generate other solutions from
the formal solution for $Q_f (\epsilon)$ with $\epsilon=1$
by Takahashi and Tanimoto \cite{Takahashi:2002ez}
because their solution is based on the identity state
which does not have any boundary.

Let us next consider energies of solutions
constructed in this method.
The computation of
$\langle B_t^\sigma ({\cal O}) |
Q_f (\epsilon) | B_t^\sigma ({\cal O}) \rangle$
can be reduced to that of
$\langle B_t ({\cal O}) |
Q_f (\epsilon) | B_t ({\cal O}) \rangle$
if the contour of the integral in $Q_f (\epsilon)$ can be deformed
across the boundary condition changing operators.
Although $Q_f (\epsilon) \cdot \sigma^{\pm}$ vanishes
in the limit $\varepsilon \to 0$,
there is a potential subtlety coming from the existence
of $\sigma^{\mp}$ from the other state.
In a coordinate
where the open-string endpoint is located at the origin,
the quantity in question is given by
\begin{eqnarray}
  &&\hskip -2cm
   (2 \varepsilon)^{2h} \, \sigma^{-} (-\varepsilon) \,
  Q_f (\epsilon) \cdot \sigma^{+} (\varepsilon)
\nonumber \\
  &=& (2 \varepsilon)^{2h} \, \sigma^{-} (-\varepsilon) \,
  \Bigl\{ h \, f'(\varepsilon) \,
  c \sigma^{+} (\varepsilon)
  + f(\varepsilon)
  \left[ h (\partial c) \sigma^{+} (\varepsilon)
  + c \partial \sigma^{+} (\varepsilon) \right] \Bigr\} \,.
\label{sigma-Q_f-sigma}
\end{eqnarray}
The leading term of the OPE
$\sigma^{-} (-\varepsilon) \,
\partial \sigma^{+} (\varepsilon)$
is of $O(\varepsilon^{-2 h-1})$ and is more singular
than that of
$\sigma^{-} (-\varepsilon) \, \sigma^{+} (\varepsilon)$,
but the quantity (\ref{sigma-Q_f-sigma}) still vanishes
in the limit $\varepsilon \to 0$
because $f(\varepsilon)$ is of $O(\varepsilon^2)$.
Similarly, we can also show that
$(2 \varepsilon)^{2h} \, \left[ \, Q_f (\epsilon) \cdot
\sigma^{-} (-\varepsilon) \, \right] \, \sigma^{+} (\varepsilon)$
vanishes in the limit $\varepsilon \to 0$.
Therefore, we can safely deform
the contour of the integral in $Q_f (\epsilon)$
across any of the boundary condition changing operators.
Then, each of the two pairs of adjacent $\sigma^+$ and $\sigma^-$
can be replaced by $(2 \varepsilon)^{-2h}$, and the two factors
of $(2 \varepsilon)^{2h}$ from the normalization of
$| B_t^\sigma ({\cal O}) \rangle$ are canceled.
Similarly, each of the three pairs
of adjacent $\sigma^+$ and $\sigma^-$
in $\langle B_t^\sigma ({\cal O}) |
B_t^\sigma ({\cal O}) \ast B_t^\sigma ({\cal O}) \rangle$
can be replaced by $(2 \varepsilon)^{-2h}$, and the three factors
of $(2 \varepsilon)^{2h}$ from the normalization of
$| B_t^\sigma ({\cal O}) \rangle$ are canceled.
In both cases, the remaining computation is the same
except that the boundary condition has been changed.
Since the operators ${\cal O}$ and $Q_f (\epsilon)$ consist of
$T^m$, $b$, and $c$, either of
$\langle B_t^\sigma ({\cal O}) |
Q_f (\epsilon) | B_t^\sigma ({\cal O}) \rangle$
and $\langle B_t^\sigma ({\cal O}) |
B_t^\sigma ({\cal O}) \ast B_t^\sigma ({\cal O}) \rangle$
is a correlation function of operators $T^m$, $b$, and $c$.
The ghost part of the correlation is obviously independent
of the boundary condition.
The matter part is made of correlation functions of $T^m$,
which are completely determined by the conformal symmetry
except for the overall normalization
given by the disk partition function
${\cal Z}_\sigma = \vev{1}$ with the boundary condition
associated with $\sigma^{\pm}$.
Therefore, we find
\begin{equation}
  \frac{\langle B_t^\sigma ({\cal O}) |
  Q_f (\epsilon) | B_t^\sigma ({\cal O}) \rangle}
  {\langle B_t^{\sigma'} ({\cal O}) |
  Q_f (\epsilon) | B_t^{\sigma'} ({\cal O}) \rangle}
  = \frac{\langle B_t^\sigma ({\cal O}) |
  B_t^\sigma ({\cal O}) \ast B_t^\sigma ({\cal O}) \rangle}
  {\langle B_t^{\sigma'} ({\cal O}) |
  B_t^{\sigma'} ({\cal O}) \ast B_t^{\sigma'} ({\cal O}) \rangle}
  = \frac{{\cal Z}_\sigma}{{\cal Z}_{\sigma'}}
\label{energy-ratios}
\end{equation}
for any pair of boundary conditions
associated with $\sigma^{\pm}$ and $\sigma'^{\pm}$.
It follows from (\ref{energy-ratios}) that
ratios of energies of solutions
are given by ratios of disk partition functions.
Since the disk partition function
is proportional to the corresponding D-brane tension
\cite{Callan:1995xx, DiVecchia:1997pr, Elitzur:1998va,
Harvey:1999gq, deAlwis:2001hi},
ratios of D-brane tensions are correctly reproduced
just as in the case of  \cite{Rastelli:2001vb}.
In our case, however, their assumption of the factorization
into the  matter and ghost sectors has been relaxed.
Again, this argument holds
for any solution which consists of a surface state
with insertions of $T^m$, $b$, and $c$
when the kinetic operator
is made of $T^m$, $b$, and $c$ and its action
on the boundary condition changing operators vanishes.
Furthermore, the derivation of (\ref{energy-ratios})
did not require the equation of motion to be satisfied
so that we can use (\ref{energy-ratios})
even for an approximate solution
such as the level-truncation solution or the one
based on the regulated butterfly state.

\section{Discussion}
\label{section-5}

The conjecture in \cite{Rastelli:2000hv}
that the kinetic operator ${\cal Q}$ can be made
purely of ghost fields was motivated by universality.
Namely, the open string field theory action at the tachyon vacuum
should be independent of the matter sector,
although a reference matter boundary CFT with $c=26$
is still necessary to formulate string field theory.
The kinetic operator $Q_f (\epsilon)$ apparently depends
on the matter boundary CFT
and the resulting string field theory
may not seem to be universal.
However, when the kinetic operator
is given by $Q_f (\epsilon)$,
the open-string end points do not propagate
in the Siegel gauge
because of the zeros of $f(\xi)$ at the open-string end points
\cite{Drukker:2002ct, Drukker:2003hh,Zeze:2004yh}.
It is expected that the resulting Riemann surfaces
are effectively closed
and the amplitudes they express could be
independent of the open-string boundary condition in this way.

On the other hand, the open-string midpoint does propagate
in the Siegel gauge. This resolves the following problematic
singular nature of the midpoint $c$-ghost insertion.
Consider the one-loop effective action
of vacuum string field theory
evaluated at the D25-brane solution.
It should reproduce the ordinary amplitudes
integrated over the cylinder modulus.
However, when the kinetic operator
is made purely of ghost fields,
not only the end points but also the whole open string
does not propagate.
Then whatever state we may take for the D25-brane solution,
it seems that only degenerated cylinders can be generated
because the open-string midpoint does not propagate \cite{OOO}.
The kinetic operator $Q_f (\epsilon)$ seems to avoid this problem
without breaking universality.

The argument in section \ref{section-4} only requires that
the kinetic operator should consist of $b$, $c$, and $T^m$
and that the action of the kinetic operator
on boundary condition changing operators at the open-string
end points gives vanishing contributions.
The kinetic operator $Q_f (\epsilon)$ considered in this paper
is one particular example, and there will be
other operators satisfying these requirements.
For example, we can choose a different function $f(\xi)$
for $Q_f$.\footnote
{
Kinetic operators with different choices of $f(\xi)$ were studied
in \cite{Kishimoto:2002xi} and \cite{Igarashi:2005wh}.
}
It is an important open question
if they are equivalent or not.
There may also be operators satisfying the requirements
other than the ones constructed by Takahashi and Tanimoto.
However, it is very nontrivial
to construct consistent kinetic operators
having the nilpotency and the derivation property
out of $b$, $c$, and $T^m$ in other ways.
As far as we can see there are no obvious problems with 
the kinetic operators $Q_f (\epsilon)$.
These operators could be
exact kinetic operators at the tachyon vacuum.

Once the singular pure-ghost theory of Gaiotto, Rastelli, 
Sen and Zwiebach is related to a regular theory
by field redefinition, it is not surprising
that the D25-brane solution has finite energy density.
A nontrivial question is if the energy density ${\cal E}$
and the on-shell three-tachyon coupling constant $g_T$
on a D25-brane are related by the formula
${\cal E} = 1/(2 \pi^2 \alpha'^3 g_T^2)$.
This relation, together with the correct open-string mass spectrum,
can be derived
if we assume the matter-ghost factorization \cite{Okawa:2002pd}.
It is an important open problem
if this relation and the mass spectrum can be derived
for the theory with the operator $Q_f (\epsilon)$.

\section*{Acknowledgments}
Y.O. would like to thank Wati Taylor and Barton Zwiebach
for useful discussions.
N.D. would like to thank the hospitality of the Center for Theoretical 
Physics at MIT and of the 
Aspen Center for Physics,
where part of this work was done.
The work of Y.O. is supported in part by funds
provided by the U.S. Department of Energy (D.O.E.)
under cooperative research agreement DE-FC02-94ER40818.


\appendix
\renewcommand{\thesection}{Appendix \Alph{section}.}
\renewcommand{\theequation}{\Alph{section}.\arabic{equation}}

\section{Conformal field theory formulation of string field theory}

In the CFT formulation of string field theory
\cite{LeClair:1988sp, LeClair:1988sj},
an open string field is represented
as a wave functional obtained by a path integral
over a certain region in a Riemann surface.
For example, a state $\ket{\phi}$ in the Fock space
can be represented as a wave functional on the arc
$| \xi |=1$ in the upper-half complex plane of $\xi$
by path-integrating over the interior of
the upper half of the unit disk $| \xi | < 1$
with the corresponding operator $\phi (0)$
inserted at the origin
and with the boundary condition of the open string
imposed on the part of the real axis $-1 \le \xi \le 1$.
A more general class of states
such as the regulated butterfly state
can be defined by a path integral over a different region of
a Riemann surface with a boundary
and with possible operator insertions.
When we parametrize the open string on the arc
as $\xi = e^{i \theta}$ with $0 \le \theta \le \pi$,
we refer to the region $\pi/2 \le \theta \le \pi$
as the left half of the open string,
and to the region $0 \le \theta \le \pi/2$
as the right half of the open string.
We also refer to the point $\theta=\pi/2$
as the open-string midpoint.

We use the standard definitions \cite{Witten:1985cc} of
the inner product $\vev{\phi_1 | \phi_2}$
and the star product $\ket{\phi_1 \ast \phi_2}$.
The state $\ket{\phi_1 \ast \phi_2}$ is defined by
gluing together
the right half of the open string of $\ket{\phi_1}$
and the left half of the open string of $\ket{\phi_2}$.
Gluing can be performed by conformal transformations
which map the two regions to be glued together
into the same region.
The inner product $\vev{\phi_1 | \phi_2}$ is defined
by gluing the left and right halves of
the open string of $\ket{\phi_1 \ast \phi_2}$.

We use the doubling trick throughout the paper.
For example, $bc$ ghosts on the upper-half plane
are extended to the lower-half plane by
$c (\bar{z}) = \tilde{c} (z)$ and $b (\bar{z}) = \tilde{b} (z)$.
The normalization of correlation functions is given by
\begin{equation}
  \vev{ c(z_1) \, c(z_2) \, c(z_3) }
  = ( z_1 - z_2 )( z_1 - z_3 )( z_2 - z_3 ) \int d^{26} x \,.
\label{three-c}
\end{equation}
In this paper, we only consider correlation functions
which are independent of space-time coordinates so that
the space-time volume always factors out.
We use the subscript $density$ to denote
a quantity divided by the volume factor of space-time.
For example, (\ref{three-c}) is written as
\begin{equation}
  \vev{ c(z_1) \, c(z_2) \, c(z_3) }_{density}
  = ( z_1 - z_2 )( z_1 - z_3 )( z_2 - z_3 ) \,.
\end{equation}

The normalization of a state $\ket{\phi}$ in the Fock space
is fixed by the condition
that the $SL(2,R)$-invariant vacuum $\ket{0}$
corresponds to the identity operator. From the normalization
of correlation functions (\ref{three-c})
and the standard mode expansion of $c$ on the unit circle
given by
\begin{equation}
  c_n = \oint \frac{dz}{2 \pi i} \, z^{n-2} \, c(z) \,,
\end{equation}
the normalization of the inner product is then fixed as follows:
\begin{equation}
  \vev{ 0 | c_{-1} c_{0} c_{1} | 0 }_{density} = 1 \,.
\end{equation}

\section{Solution
at the next-to-leading order
using the butterfly state}

Our ansatz for the solution at the next-to-leading order is
\begin{equation}
  | \Psi^{(2)} \rangle = \frac{x}{\sqrt{1-t}} \ket{B_t(c)}
  + \sqrt{1-t} \left[ u \, | B_t(\partial^2 c) \rangle
  + v \, \ket{ B_t(c T^m) }
  + w \, \ket{ B_t(\no{ bc \partial c}) } \right],
\label{Psi^(2)}
\end{equation}
where $T^m$ is the matter part of the energy-momentum tensor
and $x$, $u$, $v$, and $w$ are parameters to be determined.

The quantity
$\langle \phi | Q_f (\epsilon) | \Psi^{(2)} \rangle$
can be computed as in the case of
$\langle \phi | Q_f (\epsilon) | \Psi^{(0)} \rangle$.
We use the following OPE's
up to order of operators with dimension one:
\begin{eqnarray}
  j_B (z) \,  c (0)
  &=& \frac{1}{z} \, c \partial c (0)
  - \frac{1}{2} \, c \partial^2 c (0)
  + z \, \left[ \, -c\partial^3c (0)
  +\frac{1}{2} \partial c \partial^2 c (0)
  -c\partial c T^m (0) \, \right]
  + O(z^2) \,,
\\
  j_B (z) \, \partial^2 c (0)
  &=& \frac{2}{z^3} \, c \partial c (0)
  + \frac{2}{z^2} \, c \partial^2 c (0)
  +\frac{1}{z} \, \biggl[ \, c\partial^3 c (0)
  +\partial c \partial^2 c (0) \, \biggr]
  + O(z^0) \,,
\\
  j_B (z) \, c T^m (0)
  &=& -\frac{13}{z^3} \, c \partial c (0)
  -\frac{13}{2} \frac{1}{z^2} \, c \partial^2 c (0)
  +\frac{1}{z} \, \left[ \, -\frac{13}{6} \, c \partial^3 c (0)
  - c \partial c T^m (0) \, \right]
  + O(z^0) \,,
\\
  j_B (z) \, \no{ bc \partial c } (0)
  &=& \frac{6}{z^3} c \partial c (0)
  +\frac{3}{2} \frac{1}{z^2} \, c \partial^2 c (0)
  +\frac{1}{z} \, \left[ \, \frac{2}{3} \, c \partial^3 c (0)
  -\frac{3}{2} \, \partial c \partial^2 c (0)
  + c \partial c T^m (0) \, \right]
  + O(z^0) \,, \qquad
\\
  c (z) \, c (0)
  &=& - z\, c \partial c (0)
  - \frac{z^2}{2}\, c \partial^2 c (0)
  - \frac{z^3}{6}\, c \partial^3 c (0)
  + O(z^4) \,,
\label{c-c-OPE}
\\
  c (z) \, \partial^2 c (0)
  &=&  c \partial^2 c (0)
  + z\, \partial c \partial^2 c (0)
  + O(z^3) \,,
\label{c-d^2c-OPE}
\\
  c (z) \, c T^m (0)
  &=&   - z \, c \partial c T^m (0)
  + O(z^2) \,,
\label{c-cT-OPE}
\\
  c (z) \, \no{ bc \partial c } (0)
  &=& \frac{1}{z} \, c \partial c (0)
  + O(z^2) \,.
\label{c-bcdc-OPE}
\end{eqnarray}
We also use the OPE's with $c(z)$
replaced by $\partial c(z)$ or by $\partial^2 c(z)$
in (\ref{c-c-OPE}), (\ref{c-d^2c-OPE}),
(\ref{c-cT-OPE}), and (\ref{c-bcdc-OPE}),
which are easily derived by taking $z$ derivatives
of the OPE's with $c(z)$.
We also need the term of $O(z^2)$ in (\ref{f(z)-expansion}),
which is given by
\begin{eqnarray}
  f(z) &=& - \frac{\epsilon^2 \, (2 + p^2 + \epsilon \, p^2)^2}
  {p^2 \, (1 + \epsilon)^2 \,
  (p^2 - 4 \, \epsilon - \epsilon^2 \, p^2)}
  -\frac{4 \, \epsilon^2 \, (2 + p^2 + \epsilon \, p^2) \, 
  (p^2 - 2 \, \epsilon + \epsilon \, p^2)}
  {p^4 \, (1 + \epsilon)^2 \,
  (p^2 - 4 \, \epsilon - \epsilon^2 \, p^2)^2} \, z^2
  + O(z^4)
\nonumber \\
  &=& \left[ \, - \frac{4 \, \epsilon^2}{p^2 \, (p^2 - 4 \, \epsilon)}
  + O(\epsilon) \, \right]
  + \left[ \, -\frac{8 \, \epsilon^2 \, (p^2 - 2 \, \epsilon)}
  {p^4 \, (p^2 - 4 \, \epsilon)^2} + O(\epsilon^0)
  \, \right] \, z^2 + O(z^4) \,.
\end{eqnarray}
The final form of
$\langle \phi | Q_f (\epsilon) | \Psi^{(2)} \rangle$
is then given by
\begin{eqnarray}
  \langle \phi | Q_f (\epsilon) | \Psi^{(2)} \rangle
  &=& \left\{ \left( \frac{\eta}{2 \, (1+\eta)} - \frac{1}{\eta}
    \right) x
    -\frac{\eta}{4 \, (1+\eta)^2}
    \left(2 \, u -13 \, v +6 \, w \right)
    -\frac{w}{2 \, \eta} + O(\epsilon) \right\}
    \frac{\vev{ \phi | B_t(c \partial c)}}{\sqrt{1-t}}
\nonumber \\ && {}
  + \left\{ \frac{x}{3 \, \eta} + \frac{\eta}{2 \, (1+\eta)}
  \left( u -\frac{13}{6} \, v + \frac{2}{3} \, w \right) \right\}
  \sqrt{1-t} \, \langle \phi | B_t(c \partial^3 c) \rangle
\nonumber \\ && {}
  + \left\{ -\frac{\eta \, x}{2} +\frac{\eta}{2 \, (1+\eta)}
  \left( u -\frac{3}{2} \, w \right) +\frac{u}{\eta} \right\}
  \sqrt{1-t} \, \langle \phi | B_t(\partial c \partial^2 c) \rangle
\nonumber \\ && {}
  + \left\{ \eta \, x +\frac{\eta}{2 \, (1+\eta)}
  ( -v+w ) -\frac{v}{\eta} \right\}
  \sqrt{1-t} \, \vev{ \phi | B_t(c \partial c T^m) }
  + O(\epsilon^{3/2}) \,.
\label{phi-Q_f-Psi-2}
\end{eqnarray}

The quantity $\langle \phi | \Psi^{(2)} \ast \Psi^{(2)} \rangle$
has been computed in \cite{Okawa:2003zc}.
The leading part of the equation of motion
$\langle \phi | Q_f | \Psi^{(2)} \rangle
+ \langle \phi | \Psi^{(2)} \ast \Psi^{(2)} \rangle = 0$,
which is proportional to $\vev{\phi | B(c \partial c)}$
and is of $O(1/\sqrt{1-t})$, is given by
\begin{equation}
  \left( \frac{\eta}{2 \, (1+\eta)} - \frac{1}{\eta} \right) x
  -\frac{\eta}{4 \, (1+\eta)^2}
  \left(2 \, u -13 \, v +6 \, w \right)
  -\frac{w}{2 \, \eta} + {\cal F} = 0 \,,
\label{equation-1}
\end{equation}
where
\begin{eqnarray}
  {\cal F} &=& \frac{3^{3/4}}{\sqrt{2}} \left(
  \frac{8\,x^2}{3}
  + \frac{314\,x\,u}{81}
  + \frac{2585\,u^2}{2187}
  - \frac{247\,x\,v}{27}
  - \frac{38779\,u\,v}{5832}
  + \frac{35243\,v^2}{3888} \right.
\nonumber \\
  && \left. {}
  + \frac{7\,x\,w}{3}
  + \frac{919\,u\,w}{648}
  - \frac{1729\,v\,w}{432}
  + \frac{w^2}{2} \right) \,.
\end{eqnarray}
The three other equations, of $O(\sqrt{1-t})$,
for the coefficients in front of
$\langle \phi | B(c \partial^3 c) \rangle$,
$\langle \phi | B(\partial c \partial^2 c) \rangle$,
and $\vev{ \phi | B(c \partial c T^m) }$ are
\begin{eqnarray}
  && \frac{x}{3 \, \eta} + \frac{\eta}{2 \, (1+\eta)}
  \left( u -\frac{13}{6} \, v + \frac{2}{3} \, w \right)
  - \frac{2}{3}\left(1-\frac{1}{\sqrt{3}}\right){\cal F}
\nonumber \\
  && \quad {}+ \frac{\sqrt{2}}{3^{3/4}} \left(
  \frac{4\,x^2}{9}
  + \frac{517\,x\,u}{243}
  + \frac{22385\,u^2}{13122}
  - \frac{247\,x\,v}{162}
  - \frac{127699\,u\,v}{34992}
  + \frac{35243\,v^2}{23328} \right.
\nonumber \\
  && \quad \left. {}
  + \frac{5\,x\,w}{6}
  + \frac{965\,u\,w}{1296}
  - \frac{1235\,v\,w}{864}
  + \frac{w^2}{4}\right) = 0 \,,
\label{equation-2}
\\
  &&-\frac{\eta \, x}{2} +\frac{\eta}{2 \, (1+\eta)}
  \left( u -\frac{3}{2} \, w \right) +\frac{u}{\eta}
  +\frac{3}{2}\left(1-\frac{1}{\sqrt{3}}\right){\cal F}
\nonumber \\
  && \quad {}+ \frac{\sqrt{2}}{3^{3/4}} \left(
  - \frac{4\,x^2}{3}
  - \frac{193\,x\,u}{81}
  - \frac{3431\,u^2}{4374}
  + \frac{247\,x\,v}{54}
  + \frac{47671\,u\,v}{11664}
  - \frac{35243\,v^2}{7776} \right.
\nonumber \\
  && \quad \left. {}
  - \frac{9\,x\,w}{2}
  - \frac{559\,u\,w}{144}
  + \frac{247\,v\,w}{32}
  - \frac{23\,w^2}{12}\right) = 0 \,,
\label{equation-3}
\\
  && \eta \, x +\frac{\eta}{2 \, (1+\eta)} ( -v+w ) -\frac{v}{\eta}
  -\left(1-\frac{1}{\sqrt{3}}\right){\cal F}
\nonumber \\
  && \quad {}+ \frac{\sqrt{2}}{3^{3/4}} \left(
  4\,x\,v
  + \frac{157\,u\,v}{54}
  - \frac{55\,v^2}{9}
  + \frac{7\,v\,w}{4}\right) = 0 \,.
\label{equation-4}
\end{eqnarray}
There are also terms of $O(\sqrt{1-t})$
which are proportional to $\vev{ \phi | B(c \partial c) }$.
The equation coming from $\vev{ \phi | B(c \partial c) }$
at this subleading order can be easily satisfied
by introducing a subleading part of $x$.
We will not compute it because
it does not contribute to the energy
density of the solution in the limit $t \to 1$.

We have the four equations (\ref{equation-1}), (\ref{equation-2}),
(\ref{equation-3}), and (\ref{equation-4}) to solve
for the four variables $x$, $u$, $v$, and $w$ for a given $\eta$.
We could not solve them analytically,
but it is possible to find
numerical solutions using {\it Mathematica} for a given $\eta$.
The solutions depend on the value of $\eta$, and some branches of
real-valued solutions exist for only a limited range of that parameter.

We then need to compute
$\langle \Psi^{(2)} | Q_f (\epsilon) | \Psi^{(2)} \rangle$
in evaluating ${\cal R} [ \Psi^{(2)} ]$
and ${\cal V} [ \Psi^{(2)} ]$ for the solutions.
As in the case of the computation at the leading order,
the quantity
$\langle \Psi^{(2)} | Q_f (\epsilon) | \Psi^{(2)} \rangle$
can be related to
$\langle \widetilde{\Psi}^{(2)} |
Q_f (\eta) | \widetilde{\Psi}^{(2)} \rangle$
computed for level truncation in subsection \ref{subsection-3.1}.
The conformal transformations of the operators
$\partial^2 c$, $c T^m$, and $\no{ bc \partial c}$
in $| \widetilde{\Psi}^{(2)} \rangle$
to the $z$ coordinate related to $\xi$ by (\ref{xi-z})
have been computed in~\cite{Okawa:2003zc},
and their further transformations
to the $w$ coordinate (\ref{w-coordinate}) are easily derived.
The upshot is that the quantity
$\langle \Psi^{(2)} | Q_f (\epsilon) | \Psi^{(2)} \rangle$
can be obtained from
$\langle \widetilde{\Psi}^{(2)} |
Q_f (\eta) | \widetilde{\Psi}^{(2)} \rangle$
by the following replacements:
\begin{eqnarray}
  && x \to \sqrt{\frac{1-t^4}{1-t}} \, x
  + \sqrt{\frac{1-t}{1-t^4}} \left( \, 3 \, t^2 \, u
  - \frac{13 \, t^2}{2} \, v + 2 \, t^2 \, w \, \right) \,,
\nonumber \\
  && u \to \sqrt{\frac{1-t}{1-t^4}} \, u \,, \qquad
  v \to \sqrt{\frac{1-t}{1-t^4}} \, v \,, \qquad
  w \to \sqrt{\frac{1-t}{1-t^4}} \, w \,.
\end{eqnarray}
After the replacements and taking the limit $t \to 1$, we find
\begin{eqnarray}
  &&\hskip-2cm \lim_{t \to 1}
  \langle \Psi^{(2)} | Q_f (\epsilon) | \Psi^{(2)} \rangle_{density}
\nonumber \\
  &=& \frac{\eta}{1+ \eta} \left[
  - \left( 2\,x + \frac{3}{2}\, u - \frac{13}{4} \,
        v + \, w \right)^2
  -u^2 + \frac{13}{4} \, v^2 + \, w^2
  +3 \, u \, w - \frac{13}{2} \, v \, w
  \right]
\nonumber \\ && 
  {}- \frac{2 \, \eta}{(1 + \eta)^2}
  \left( 2\,x + \frac{3}{2}\, u - \frac{13}{4} \,
        v + \, w \right)
  ( - u + \frac{13}{2} \, v - 3 \, w )
\nonumber \\ && 
  {}+ \frac{1 + 3 \, \eta^2}{2 \, \eta} \left[
  \left( 2\,x + \frac{3}{2}\, u - \frac{13}{4} \,
        v + \, w \right)^2
  -  u^2 + \frac{13}{4} \, v^2 \right]
\nonumber \\ && 
  {}+ \frac{1 - 3 \, \eta^2}{\eta}
  \left( 2\,x + \frac{3}{2}\, u - \frac{13}{4} \,
        v + \, w \right) \frac{w}{2} \,.
\end{eqnarray}
The quantity
$\langle \Psi^{(2)} | \Psi^{(2)} \ast \Psi^{(2)} \rangle$
has been computed in \cite{Okawa:2003zc}.
In the limit $t \to 1$, it is given by
\begin{eqnarray}
  &&\hskip-2cm \lim_{t \to 1}
  \langle \Psi^{(2)} | \Psi^{(2)} \ast \Psi^{(2)} \rangle_{density}
  = -\frac{81\,{\sqrt{3}}\,x^3}{8}
  -\frac{927\,{\sqrt{3}}\,x^2\,u}{32}
  -\frac{3451\,{\sqrt{3}}\,x\,u^2}{128}
\nonumber \\ &&  {}
  -\frac{4205\,{\sqrt{3}}\,u^3}{512}
  +\frac{4329\,{\sqrt{3}}\,x^2\,v}{64}
  +\frac{49543\,x\,u\,v}{128\,{\sqrt{3}}}
  +\frac{1659931\,u^2\,v}{9216\,{\sqrt{3}}}
  -\frac{244673\,x\,v^2}{512\,{\sqrt{3}}}
\nonumber \\ &&  {}
  -\frac{25201319\,u\,v^2}{55296\,{\sqrt{3}}}
  +\frac{43213963\,v^3}{110592\,{\sqrt{3}}}
  -\frac{315\,{\sqrt{3}}\,x^2\,w}{16}
  -\frac{1095\,{\sqrt{3}}\,x\,u\,w}{32}
\nonumber \\ &&  {}
  -\frac{286195\,u^2\,w}{6912\,{\sqrt{3}}}
  +\frac{16835\,x\,v\,w}{64\,{\sqrt{3}}}
  +\frac{175565\,u\,v\,w}{768\,{\sqrt{3}}}
  -\frac{8563555\,v^2\,w}{27648\,{\sqrt{3}}}
\nonumber \\ &&  {}
  -\frac{403\,{\sqrt{3}}\,x\,w^2}{32}
  -\frac{104687\,u\,w^2}{3456\,{\sqrt{3}}}
  +\frac{193843\,v\,w^2}{2304\,{\sqrt{3}}}
  -\frac{169\,{\sqrt{3}}\,w^3}{64} \,.
\end{eqnarray}
We can now evaluate ${\cal R} [ \Psi^{(2)} ]$
and ${\cal V} [ \Psi^{(2)} ]$ in the limit $t \to 1$
for the numerical solutions for a given $\eta$.
The results are summarized in table \ref{ratio-tension-eta}.


\renewcommand{\baselinestretch}{0.87}


\begin{thebibliography}{10}

\bibitem{Rastelli:2000hv}
L.~Rastelli, A.~Sen and B.~Zwiebach,
``String field theory around the tachyon vacuum,''
Adv.\ Theor.\ Math.\ Phys.\  {\bf 5}, 353 (2002)
[arXiv:hep-th/0012251].

\bibitem{Rastelli:2001jb}
L.~Rastelli, A.~Sen and B.~Zwiebach,
``Classical solutions in string field theory
around the tachyon vacuum,''
Adv.\ Theor.\ Math.\ Phys.\  {\bf 5}, 393 (2002)
[arXiv:hep-th/0102112].

\bibitem{Rastelli:2001uv}
L.~Rastelli, A.~Sen and B.~Zwiebach,
``Vacuum string field theory,''
arXiv:hep-th/0106010.

\bibitem{Witten:1985cc}
E.~Witten,
``Noncommutative geometry and string field theory,''
Nucl.\ Phys.\ B {\bf 268}, 253 (1986).

\bibitem{Taylor:2003gn}
W.~Taylor and B.~Zwiebach,
``D-Branes, tachyons, and string field theory,''
arXiv:hep-th/0311017.

\bibitem{Kostelecky:2000hz}
V.~A.~Kostelecky and R.~Potting,
``Analytical construction of a nonperturbative vacuum
for the open bosonic string,''
Phys.\ Rev.\ D {\bf 63}, 046007 (2001)
[arXiv:hep-th/0008252].

\bibitem{Rastelli:2001vb}
L.~Rastelli, A.~Sen and B.~Zwiebach,
``Boundary CFT construction of D-branes
in vacuum string field theory,''
JHEP {\bf 0111}, 045 (2001)
[arXiv:hep-th/0105168].

\bibitem{Rastelli:2000iu}
L.~Rastelli and B.~Zwiebach,
``Tachyon potentials, star products and universality,''
JHEP {\bf 0109}, 038 (2001)
[arXiv:hep-th/0006240].

\bibitem{Gaiotto:2001ji}
D.~Gaiotto, L.~Rastelli, A.~Sen and B.~Zwiebach,
``Ghost structure and closed strings in vacuum string field theory,''
Adv.\ Theor.\ Math.\ Phys.\  {\bf 6}, 403 (2003)
[arXiv:hep-th/0111129].

\bibitem{Schnabl:2002ff}
M.~Schnabl,
``Anomalous reparametrizations and butterfly states
in string field  theory,''
Nucl.\ Phys.\ B {\bf 649}, 101 (2003)
[arXiv:hep-th/0202139].

\bibitem{Gaiotto:2002kf}
D.~Gaiotto, L.~Rastelli, A.~Sen and B.~Zwiebach,
``Star algebra projectors,''
JHEP {\bf 0204}, 060 (2002)
[arXiv:hep-th/0202151].

\bibitem{Okawa:2003cm}
Y.~Okawa,
``Some exact computations on the twisted butterfly state
in string field theory,''
JHEP {\bf 0401}, 066 (2004)
[arXiv:hep-th/0310264].

\bibitem{LeClair:1988sp}
A.~LeClair, M.~E.~Peskin and C.~R.~Preitschopf,
``String field theory
on the conformal plane. 1. Kinematical principles,''
Nucl.\ Phys.\ B {\bf 317}, 411 (1989).

\bibitem{LeClair:1988sj}
A.~LeClair, M.~E.~Peskin and C.~R.~Preitschopf,
``String field theory
on the conformal plane. 2. Generalized gluing,''
Nucl.\ Phys.\ B {\bf 317}, 464 (1989).

\bibitem{Mukhopadhyay:2001ey}
P.~Mukhopadhyay,
``Oscillator representation of the BCFT construction
of D-branes in vacuum string field theory,''
JHEP {\bf 0112}, 025 (2001)
[arXiv:hep-th/0110136].

\bibitem{Okuyama:2002tw}
K.~Okuyama,
``Ratio of tensions from vacuum string field theory,''
JHEP {\bf 0203}, 050 (2002)
[arXiv:hep-th/0201136].

\bibitem{Okawa:2002pd}
Y.~Okawa,
``Open string states and D-brane tension from
vacuum string field theory,''
JHEP {\bf 0207}, 003 (2002)
[arXiv:hep-th/0204012].

\bibitem{Hata:2001sq}
H.~Hata and T.~Kawano,
``Open string states around a classical solution
in vacuum string field theory,''
JHEP {\bf 0111}, 038 (2001)
[arXiv:hep-th/0108150].

\bibitem{Okawa:2003zc}
Y.~Okawa,
``Solving Witten's string field theory using the butterfly state,''
Phys.\ Rev.\ D {\bf 69}, 086001 (2004)
[arXiv:hep-th/0311115].

\bibitem{Drukker:2003hh}
N.~Drukker,
``On different actions for the vacuum
of bosonic string field theory,''
JHEP {\bf 0308}, 017 (2003)
[arXiv:hep-th/0301079].

\bibitem{Takahashi:2002ez}
T.~Takahashi and S.~Tanimoto,
``Marginal and scalar solutions in cubic open string field theory,''
JHEP {\bf 0203}, 033 (2002)
[hep-th/0202133].

\bibitem{Bonora:2003cc}
L.~Bonora, C.~Maccaferri and P.~Prester,
``Dressed sliver solutions in vacuum string field theory,''
JHEP {\bf 0401}, 038 (2004)
[arXiv:hep-th/0311198].

\bibitem{Bonora:2004qj}
L.~Bonora, C.~Maccaferri and P.~Prester,
``The perturbative spectrum of the dressed sliver,''
Phys.\ Rev.\ D {\bf 71}, 026003 (2005)
[arXiv:hep-th/0404154].

\bibitem{Kishimoto:2002xi}
I.~Kishimoto and T.~Takahashi,
``Open string field theory around universal solutions,''
Prog.\ Theor.\ Phys.\  {\bf 108}, 591 (2002)
[hep-th/0205275].

\bibitem{Ellwood:2001py}
I.~Ellwood and W.~Taylor,
``Open string field theory without open strings,''
Phys.\ Lett.\ B {\bf 512}, 181 (2001)
[arXiv:hep-th/0103085].

\bibitem{Giusto:2003wc}
S.~Giusto and C.~Imbimbo,
``Physical states at the tachyonic vacuum
of open string field theory,''
Nucl.\ Phys.\ B {\bf 677}, 52 (2004)
[arXiv:hep-th/0309164].

\bibitem{Takahashi:2003xe}
T.~Takahashi and S.~Zeze,
``Gauge fixing and scattering amplitudes in string field theory
around universal solutions,''
Prog.\ Theor.\ Phys.\  {\bf 110}, 159 (2003)
[arXiv:hep-th/0304261].

\bibitem{Bars:2003cr}
I.~Bars, I.~Kishimoto and Y.~Matsuo,
``Analytic study of nonperturbative solutions
in open string field  theory,''
Phys.\ Rev.\ D {\bf 67}, 126007 (2003)
[arXiv:hep-th/0302151].

\bibitem{Sen:1999nx}
A.~Sen and B.~Zwiebach,
``Tachyon condensation in string field theory,''
JHEP {\bf 0003}, 002 (2000)
[arXiv:hep-th/9912249].

\bibitem{Moeller:2000xv}
N.~Moeller and W.~Taylor,
``Level truncation and the tachyon
in open bosonic string field theory,''
Nucl.\ Phys.\ B {\bf 583}, 105 (2000)
[arXiv:hep-th/0002237].

\bibitem{Taylor:2002fy}
W.~Taylor,
``A perturbative analysis of tachyon condensation,''
JHEP {\bf 0303}, 029 (2003)
[arXiv:hep-th/0208149].

\bibitem{Gaiotto:2002wy}
D.~Gaiotto and L.~Rastelli,
``Experimental string field theory,''
JHEP {\bf 0308}, 048 (2003)
[arXiv:hep-th/0211012].

\bibitem{Ellwood:2001ne}
I.~Ellwood and W.~Taylor,
``Gauge invariance and tachyon condensation
in open string field theory,''
arXiv:hep-th/0105156.

\bibitem{Takahashi:2003pp}
T.~Takahashi,
``Tachyon condensation and universal solutions
in string field theory,''
Nucl.\ Phys.\ B {\bf 670}, 161 (2003)
[arXiv:hep-th/0302182].

\bibitem{Yang:2004xz}
H.~Yang,
``Solving Witten's SFT by insertion of operators on projectors,''
JHEP {\bf 0409}, 002 (2004)
[arXiv:hep-th/0406023].

\bibitem{Callan:1995xx}
C.~G.~.~Callan and I.~R.~Klebanov,
``D-brane boundary state dynamics,''
Nucl.\ Phys.\ B {\bf 465}, 473 (1996)
[arXiv:hep-th/9511173].

\bibitem{DiVecchia:1997pr}
P.~Di Vecchia, M.~Frau, I.~Pesando, S.~Sciuto, A.~Lerda and R.~Russo,
``Classical p-branes from boundary state,''
Nucl.\ Phys.\ B {\bf 507}, 259 (1997)
[arXiv:hep-th/9707068].

\bibitem{Elitzur:1998va}
S.~Elitzur, E.~Rabinovici and G.~Sarkissian,
``On least action D-branes,''
Nucl.\ Phys.\ B {\bf 541}, 246 (1999)
[arXiv:hep-th/9807161].

\bibitem{Harvey:1999gq}
J.~A.~Harvey, S.~Kachru, G.~W.~Moore and E.~Silverstein,
``Tension is dimension,''
JHEP {\bf 0003}, 001 (2000)
[arXiv:hep-th/9909072].

\bibitem{deAlwis:2001hi}
S.~P.~de Alwis,
``Boundary string field theory the boundary state formalism
and D-brane tension,''
Phys.\ Lett.\ B {\bf 505}, 215 (2001)
[arXiv:hep-th/0101200].

\bibitem{Drukker:2002ct}
N.~Drukker,
``Closed string amplitudes from
gauge fixed string field theory,''
Phys.\ Rev.\ D {\bf 67}, 126004 (2003)
[arXiv:hep-th/0207266].

\bibitem{Zeze:2004yh}
S.~Zeze,
``Worldsheet geometry of classical solutions in string field theory,''
Prog.\ Theor.\ Phys.\  {\bf 112}, 863 (2004)
[arXiv:hep-th/0405097].

\bibitem{OOO}
Y.~Okawa, T.~Okuda and H.~Ooguri, unpublished.

\bibitem{Igarashi:2005wh}
Y.~Igarashi, K.~Itoh, F.~Katsumata, T.~Takahashi and S.~Zeze,
``Classical solutions and order of zeros
in open string field theory,''
arXiv:hep-th/0502042.

\end{thebibliography}
\begingroup\raggedright\endgroup
\end{document}